\documentclass[
    aps,
    prd,
    twocolumn,
    superscriptaddress,
    nofootinbib
]{revtex4-2}

\usepackage{amsmath,amssymb,amsfonts,bbm,bm}
\usepackage{slashed}

\usepackage{tocloft} % well-formatted table of contents

\usepackage{graphicx,color}
\usepackage[colorlinks=true, linkcolor=blue, citecolor=blue, linktoc=all]{hyperref}

\usepackage[usenames,dvipsnames]{xcolor}

\usepackage{tikz-cd}
\usepackage{pict2e}
\usepackage{physics}
\usepackage{comment} 
\usepackage[makeroom]{cancel}

\newcommand{\beq}{\begin{eqnarray}}
\newcommand{\eeq}{\end{eqnarray}}
\newcommand{\beqn}{\begin{eqnarray}}
\newcommand{\eeqn}{\end{eqnarray}}
\newcommand{\bea}{\begin{eqnarray}}
\newcommand{\eea}{\end{eqnarray}}
\newcommand{\be}{\begin{equation}}
\newcommand{\ee}{\end{equation}}

\usepackage[normalem]{ulem}

\usepackage{mathrsfs,mathtools}
\renewcommand\mathbb[1]{\mathbbm{#1}}

\makeatletter
\DeclareRobustCommand{\loplus}{\mathbin{\mathpalette\dog@lsemi{+}}}
\DeclareRobustCommand{\lotimes}{\mathbin{\mathpalette\dog@lsemi{\times}}}
\DeclareRobustCommand{\roplus}{\mathbin{\mathpalette\dog@rsemi{+}}}
\DeclareRobustCommand{\rotimes}{\mathbin{\mathpalette\dog@rsemi{\times}}}

\newcommand{\dog@rsemi}[2]{\dog@semi{#1}{#2}{-90,90}}
\newcommand{\dog@lsemi}[2]{\dog@semi{#1}{#2}{270,90}}
\newcommand{\dog@semi}[3]{%
  \begingroup
  \sbox\z@{$\m@th#1#2$}%
  \setlength{\unitlength}{\dimexpr\ht\z@+\dp\z@\relax}%
  \makebox[\wd\z@]{\raisebox{-\dp\z@}{%
    \begin{picture}(1,1)
    \linethickness{\variable@rule{#1}}
    \roundcap
    \put(0.5,0.5){\makebox(0,0){\raisebox{\dp\z@}{$\m@th#1#2$}}}
    \put(0.5,0.5){\arc[#3]{0.5}}
    \end{picture}%
  }}%
  \endgroup
}
\newcommand{\variable@rule}[1]{%
  \fontdimen8  
  \ifx#1\displaystyle\textfont3\else
    \ifx#1\textstyle\textfont3\else
      \ifx#1\scriptstyle\scriptfont3\else
        \scriptscriptfont3\relax
  \fi\fi\fi
}
\DeclareRobustCommand{\loplus}{\mathbin{\mathpalette\dog@lsemi{+}}}

\usepackage{amsthm}

\renewcommand{\op}[1]{\boldsymbol{#1}}

\newcommand{\thistitle}{Toward a Unique Filter for the Gravitational Path Integral}

\theoremstyle{definition}
\newtheorem{theorem}{Theorem}[section]

\newtheorem{definition}{Definition}[section]
\newtheorem{lemma}{Lemma}[section]
\newtheorem{conjecture}{Conjecture}[section]

\begin{document}

\title{\thistitle}

\author{Marc S. Klinger}
\affiliation{Walter Burke Insitute for Theoretical Physics \\ California Institute of Technology, Pasadena, California 91125, USA}

%\date{\today}

\begin{abstract}
In recent work, McNamara and Wang demonstrated that the failure of factorization for a gravitational path integral (GPI) can be understood as an incompleteness of its spectrum of states by constructing a unitary quantum field theory (QFT) into which the GPI embeds. We examine this construction from an algebraic point of view, and use it to verify a conjecture of the author concerning the existence and uniqueness of a unitary completion for the GPI by means of the theory of conditional expectations. We explore how the unitary QFT can be regarded as a single coherent theory in which the quantitative signatures of ensemble averaging emerge purely from projecting to the underlying, smooth GPI. The conditional expectation relating the two theories may therefore be interpreted as a filter in the sense of Liu. The factorizing theory differs from a more standard QFT in the sense that it is \emph{reducible}, meaning the Hilbert space associated with the empty set is non-trivial. We explain how the structure of reducible QFTs are generically encoded in weak Hopf algebraic symmetries and demonstrate how the weak structure allows for generalized $\alpha$-sectors and half wormholes to coexist within a single Hilbert space.
\end{abstract}

\maketitle

% ------------------------------------------------------------------
% Main text
% ------------------------------------------------------------------

\section{Introduction}

A longstanding issue in the semiclassical formulation of gravity is the problem of factorization \cite{Witten:1999xp,Maldacena:2004rf}. This problem is closely related to that of unitarity \cite{Coleman:1988cy,Giddings:1987cg,Giddings:1988cx}, the information problem \cite{Hawking:1975vcx,Hawking:1976ra,Page:1993up,Polchinski:2016hrw}, and becomes especially pronounced when juxtaposed against the holographic principle. Holography dictates that a theory of quantum gravity should be dual to a lower dimensional quantum theory without gravity. As the latter is a genuine, unitary quantum theory, factorization is automatically expected and the failure of factorization on the gravity side poses an immediate obstruction to the holographic dictionary \cite{Liu:2025cml}. 

In recent work, a new perspective on the factorization problem has emerged highlighting the importance and subtlety of the semiclassical limit \cite{Liu:2025ikq,Kudler-Flam:2025cki,Klinger:2025tvg,Kudler-Flam:2026nzz,Klinger:2026kqj,Liu:2026fnd}. Restricting our attention to the AdS/CFT correspondence, the standard formulation of the large $N$ limit rests upon various choices. In the algebraic picture, for example, one must choose the operator topology with respect to which the limit is defined. As has been described in \cite{Gesteau:2025obm}, the standard choice of topology precludes the emergence of certain spacetime features in the large $N$ limit. A similar observation can be made in the path integral formulation, in which the gravitational path integral appears to perform an implicit `coarse graining' over gravitational microstates resulting in a smooth, macroscopic description that excludes certain `erratic' features \cite{Liu:2025ikq,Liu:2026fnd}. 

To organize these observations, Liu has proposed the notion of a large-$N$ filter \cite{Liu:2025ikq,Liu:2026fnd}. Namely, he argues first that general gravitational observables admit a decomposition
\beq \label{Obs = Sm + Err}
	A = A^{(\textrm{sm})} + A^{(\textrm{err})},
\eeq
where $A^{(\textrm{sm})}$ is a smooth, macroscopic contribution computable via the gravitational path integral and $A^{(\textrm{err})}$ is a microscopic contribution which is inaccessible to the smooth description. He then introduces a projection like operation, hereafter referred to as the filter $\mathbb{F}$, which eliminates the erratic contribution of any gravitational observable leaving behind only the smooth part
\beq \label{Sm = GPI}
	\mathbb{F}(A) = A^{(\textrm{sm})} = A_{\textrm{GPI}}. 
\eeq

Interpreting $A$ as the complete gravitational observable, for instance obtained from the holographically dual CFT, eqn. \eqref{Obs = Sm + Err} and \eqref{Sm = GPI} provide a possible resolution to the factorization problem -- the observables computed by the gravitational path integral needn't factorize as they don't have access to the complete data of the dual, unitary CFT. This point of view is strengthened by the observation that the non-factorizing pieces of the gravitational path integral can be isolated as the smooth projection of products of erratic observables. For instance, the non-factorizing part of the partition function in the disjoint union of closed manifolds $M_1$ and $M_2$ is given by
\begin{flalign}
	\mathcal{Z}^{\textrm{wormhole}}(M_1 \sqcup M_2) = \; &\mathbb{F}(Z(M_1 \sqcup M_2)) \nonumber \\
	&- \mathbb{F}(Z(M_1)) \mathbb{F}(Z(M_2)). 
\end{flalign}
Here, we have employed notation in which $\mathcal{Z}$ denotes the gravitational path integral and $Z$ the path integral of the dual CFT. 

An immediate question is whether a decomposition \eqref{Obs = Sm + Err} always exists, and, if it does exist, whether it is unique. Some progress in this direction was achieved in \cite{Klinger:2025tvg,Klinger:2026kqj}, in which the problem was reformulated from a complementary perspective. Given a description of only the smooth, generally non-factorizing gravitational observables $A^{(\textrm{sm})}$ is it possible to \emph{reconstruct} an extended, factorizing theory along with a projection back to the smooth part? To approach this question it is useful to recast the decomposition \eqref{Obs = Sm + Err} and filter \eqref{Sm = GPI} in an algebraic language. Eqn. \eqref{Obs = Sm + Err} implies the existence of an inclusion of the algebra of smooth observables $\mathscr{A}_{\textrm{sm}}$ inside of the complete algebra $\mathscr{A}_{\textrm{ext}}$ in which the `deficit algebra' can be identified with the algebra of erratic degrees of freedom. Given such an inclusion, the filter \eqref{Sm = GPI} is translated into a conditional expectation 
\beq
	E: \mathscr{A}_{\textrm{ext}} \rightarrow \mathscr{A}_{\textrm{sm}}, 
\eeq
which is the algebraic version of a projection to the smooth subalgebra.

Recasting the large-$N$ filter in the language of an inclusion is quite fruitful as it allows us to take advantage of an extensive literature on algebraic extensions \cite{Jones:1983kv,PimsnerPopa1986,Kosaki1991,Watatani1990,ENOCK1996466,Nill:1998iw,Longo:1989tt,Longo:1994zza,Bischoff:2014xea,DelVecchio:2017axj,AliAhmad:2025oli}. The most useful part of this theory for our purposes is the observation that inclusions admitting conditional expectations can be characterized \emph{intrinsically} in terms of categorical data contained in the smaller algebra, $\mathscr{A}_{\textrm{sm}}$ in our case, alone. The extended algebra is obtained by appending to $\mathscr{A}_{\textrm{sm}}$ new operators, called \emph{charged intertwiners}, which we may intepret as candidate, algebraic avatars of the erratic degrees of freedom. More technically, these operators describe (generalized) symmetries of the smooth algebra in the guise of formal objects called \emph{Q-systems} or \emph{Frobenius algebras} \cite{Longo:1989tt,Longo:1994zza,Bischoff:2014xea,DelVecchio:2017axj,AliAhmad:2025oli}. Each extension $\mathscr{A}_{\textrm{sm}} \hookrightarrow \mathscr{A}_{\textrm{ext}}^{(E)}$ can then be viewed rigorously as a \emph{gauging} of the symmetry identified by the conditional expectation $E$. The set of all possible extensions of the smooth data admitting an algebraic large-$N$ filter are therefore classified by which charged intertwiners are included, or equivalently which symmetry is gauged. 

In addition to formalizing the structure of the large-$N$ filter, the algebraic picture also makes manifest the connection between the factorization problem and two other prominent puzzles in semiclassical gravity -- the information problem and the closed universe problem \cite{Giddings:1988wv,McNamara:2020uza,Antonini:2023hdh,Antonini:2024mci,Antonini:2025ioh,Abdalla:2025gzn,Harlow:2025pvj,Higginbotham:2025clp,Sasieta:2025vck,Harlow:2026hky,VanRaamsdonk:2026tnv,Balasubramanian:2025jeu,Balasubramanian:2025hns,Balasubramanian:2025akx}. For a factorizing quantum field theory, the replica trick allows us to compute the von Neumann entropy via the partition function as
\begin{flalign}
	S(R) &= \lim_{n \rightarrow 1} \frac{1}{1-n} \Psi_R^{\otimes n}(s_n^R) \nonumber \\
	&= \lim_{n \rightarrow 1} \frac{1}{1-n} Z\bigg(\bigsqcup_{i = 1}^n M, s_n^R\bigg),
\end{flalign}
where $s_n^R$ is the n-fold swap operator. Replacing the partition function $Z$ with the gravitational path integral, we arrive at the standard formulation of the von Neumann entropy for a gravitating subregion:
\beq \label{GPI replica}
	\mathcal{H}(R) = \lim_{n \rightarrow 1} \frac{1}{1-n} \mathcal{Z}\bigg(\bigsqcup_{i = 1}^n M, s_n^R\bigg).
\eeq
However, the failure of factorization obstructs the equality between \eqref{GPI replica} and the definition of the von Neumann entropy from the replica trick which implicates the expectation value of the $n$-swap operator in an $n$-fold tensor product state. By consequence, the quantity \eqref{GPI replica} is in actuality a sum
\beq \label{GPI Entropy}
	\mathcal{H}(R) = \mathcal{S}(R) + \mathcal{H}_{\textrm{wormhole}}(R),
\eeq
where $\mathcal{S}(R)$ is the von Neumann entropy coming from the honest replica trick and $\mathcal{H}_{\textrm{wormhole}}(R)$ is \eqref{GPI replica} with $\mathcal{Z}$ is replaced by the wormhole contribution to the gravitational path integral. Both the gravitational derivation of the Page curve \cite{Penington:2019kki,Almheiri:2019qdq}, and the argument that the entropy of a closed universe is zero stem from interpreting $\mathcal{H}(R)$ as the genuine entropy of a single quantum state, and thus are called into question by eqn. \eqref{GPI Entropy}.  

Combining these observations, the existence of a large-$N$ filter was formulated in \cite{Klinger:2026kqj} in terms of the following definition:

\begin{definition}[Algebraic Filter] \label{Def Filter}
	Let $\mathcal{Z}$ be a gravitational path integral and denote by $\mathscr{A}_{\textrm{sm}}$ the assignment of an operator algebra to each co-dimension one manifold by cutting open $\mathcal{Z}$. An algebraic filter is a pair $(\mathcal{E}, \mathcal{C})$ where $\mathcal{E}: \mathscr{A}^{\mathcal{E}}_{\textrm{ext}} \rightarrow \mathscr{A}_{\textrm{sm}}$ is a conditional expectation and $\mathcal{C}: \mathscr{A}^{\mathcal{E}}_{\textrm{ext}} \rightarrow \mathscr{A}_{\textrm{sm}}$ is a quantum channel such that
	\begin{enumerate}
		\item $\mathcal{Z}^{\textrm{ext}}_{(\mathcal{E},\mathcal{C})} = \mathcal{Z} \circ \mathcal{C}$ is a factorizing path integral with associated operator algebras $\mathscr{A}^{\mathcal{E}}_{\textrm{ext}}$,
		\item The replica trick applied to the factorizing path integral $\mathcal{Z}^{\textrm{ext}}_{(\mathcal{E},\mathcal{C})}$ reproduces the gravitational path integral computation up to a contribution purely from the erratic operators:
		\begin{flalign}
			\mathcal{S}_{\textrm{ext}}(R) &= \lim_{n \rightarrow 1} \frac{1}{1-n} \mathcal{Z}^{\textrm{ext}}_{(\mathcal{E},\mathcal{C})}\bigg(\bigsqcup_{i = 1}^n M, s_n^R\bigg) \nonumber \\
			&= \mathcal{H}(R) + \mathcal{S}_{\textrm{err}}(R),
		\end{flalign}
		\item The erratic operators can be interpreted as half-wormholes \cite{Blommaert:2019wfy,Blommaert:2021gha,Garcia-Garcia:2021squ,Saad:2021rcu,Mukhametzhanov:2021nea,DiUbaldo:2023qli,Yang:2025kgs} which carry flux across Einstein-Rosen bridges connecting otherwise disjoint regions. 
	\end{enumerate}
\end{definition} 

Utilizing the recent work \cite{McNamara:2026isz} of McNamara and Wang (MW), the above definition can be more precisely formulated within the axiomatic formulation of the gravitational path integral. As we will explain, the main theorem of \cite{McNamara:2026isz} can be reinterpreted to imply the existence of a filter for a given gravitational path integral. The result is the following:

\begin{definition}[McNamara-Wang Filter] \label{thm: Main}
	Let $\mathcal{Z}$ be a gravitational path integral in the sense of \cite{Colafranceschi:2023moh}, which is finite, real, continuous, and reflection positive. Denote by $\mathscr{H}_{\mathcal{Z}}$ and $\mathscr{A}_{\mathcal{Z}}$ the Hilbert spaces and operator algebras obtained by cutting open $\mathcal{Z}$. Then, there exists a \emph{reducible} unitary quantum field theory $\mathcal{F}$, and a groupoid $\mathcal{G}$ such that
	\beq
		\mathcal{Z} = E(Z_{\mathcal{F}}), \qquad \mathscr{A}_{\mathcal{Z}} = E(\mathscr{A}_{\mathcal{F}}). 
	\eeq
	Here, $Z_{\mathcal{F}}$ is the factorizing partition function of the unitary QFT, and $\mathscr{A}_{\mathcal{F}}$ is its algebraic assignment. The map $E$ is the unique invariantizing conditional expectation under an action of $\mathcal{G}$. The operator algebras associated with the unitary quantum field theory are of the form $\mathscr{A}_{\mathcal{F}} \simeq \mathscr{A}_{\mathcal{Z}} \rtimes \hat{\mathcal{G}}$, where $\hat{\mathcal{G}}$ is the dual weak Hopf algebra to the groupoid algebra of $\mathcal{G}$. The algebra $\hat{\mathcal{G}}$ includes operators that encode both generalized $\alpha$-sectors and half Einstein-Rosen bridges. The pair $(E,\mathcal{F})$ is uniquely determined up to Morita equivalence e.g. there exists such a unitary QFT extending $\mathcal{Z}$ for any weak Hopf algebra $K$ whose representation category is Morita equivalent to that of $\mathcal{G}$.
\end{definition}

As we will address in the main text, this definition can be elevated to the level of a theorem in the case that $\mathcal{G}$ is a discrete groupoid. In the more general case in which $\mathcal{G}$ may be continuous, this definition should be regarded as a conjecture whose validity rests on an open problem in reconstruction for symmetric multitensor categories \cite[Sec. 10.3]{McNamara:2026isz}. 

The unitarity of $\mathcal{F}$ subsumes conditions (1) and (2) in Definition \ref{Def Filter}, leaving the specification of the map $\mathcal{C}$ implicit.\footnote{The role of $\mathcal{C}$ is replaced by the notion of a fiber functor in the axiomatic approach.} The direct translation between the $\alpha$-sectors and Einstein-Rosen bridges afforded by the categorical analysis of \cite{McNamara:2026isz} allows for condition (3) to be affirmed, as well, with $\mathcal{G}$ (or more precisely its dual) identified as the algebra of erratic operators. Thus, in the functorial language, $(\mathcal{E}, \mathcal{C})$ in Definition \ref{Def Filter} are encoded in the pair $(E,\mathcal{F})$ which play the role of a categorical filter. In the discrete case, Definition \ref{thm: Main} establishes the existence and uniqueness of a filtering operation derived purely from the data of a given gravitational path integral. 

An important feature of Definition \ref{thm: Main} is that the unitary QFT which completes the GPI is \emph{reducible}. In general, a QFT can be regarded as an assignment of each boundary manifold to a Hilbert space and each bulk manifold to a linear operator between the Hilbert spaces assigned to its boundaries. In a standard, irreducible, QFT the empty set is associated with the trivial Hilbert space $\mathbb{C}$, and consequently each closed manifold (which can be regarded as a bulk manifold with empty boundaries) is likewise mapped to a complex number we associate with the partition function. In a reducible QFT, the empty set is assigned to a non-trivial Hilbert space, naturally associated with the \emph{baby-universe Hilbert space}. As we will see, this essentially amounts to the fact that, in a reducible QFT, even the empty set can be cut open to reveal further structure. 

The organization of the paper is as follows. In Section~\ref{sec: Review} we provide a review of the construction of McNamara and Wang. In particular, their notion of the baby universe category associated with a given gravitational path integral proves to be precisely the tool necessary to diagnose its non-factorization. The main contribution of this work, therefore, is to interpret these findings within the context of the gravitational filter as originally proposed in \cite{Liu:2025ikq,Liu:2026fnd} and algebratized in \cite{Klinger:2026kqj}. In Section~\ref{sec: WHA}. We argue that the categorical construction of \cite{McNamara:2026isz} can be used to identify a \emph{single} QFT, as opposed to an ensemble of theories, with factorizing partition function from which the GPI can be recovered via a unique filtering operation. This QFT differs from more familiar QFTs in the sense that it is \emph{reducible}. In the final subsection, we demonstrate that the general algebraic structure of a reducible QFT can be encoded in a weak Hopf algebra. 

\section{Review of the McNamara-Wang Construction} \label{sec: Review}

In this section, we review the work \cite{McNamara:2026isz}. The main goal is to recall the axiomatic definition of a QFT \cite[Def. 4.6]{McNamara:2026isz}, the universal construction obtained from cutting open a partition function, the characterization of the failure of the universal construction to factorize \cite[Thm. 1.1]{McNamara:2026isz}, and the quantification of this failure in terms of the baby universe category introduced in \cite[Sec. 7]{McNamara:2026isz}. Along the way, we will explain how these concepts can be transcribed into the language of operator algebras, their inclusions, conditional expectations, and bimodule categories. 

\subsection{Background on Axiomatic Quantum Field Theory}

\begin{figure}[htbp]
    \centering

\begin{tikzpicture}[
    line width=0.8pt,
    every node/.style={font=\small},
    bordism/.style={draw, line width=0.9pt}
]

% ============================================================
% LEFT PANEL
% ============================================================

\begin{scope}[xshift=0cm]

    % Title
    \node[align=center] at (0,1.75)
        {$B\in\mathrm{Bord}_d$};

    % Circle
    \draw[bordism]
        (0,0) circle (1.0);

    % Label
    \node at (0,0) {$B$};

    % Formula
    \node[align=center] at (0,-1.55)
        {$B\mapsto\mathcal{F}(B)$};

\end{scope}

% ============================================================
% RIGHT PANEL
% ============================================================

\begin{scope}[xshift=5.2cm]

    % Title
    \node[align=center] at (0,1.95)
        {$N\in\mathrm{Bord}_d(B_1,B_2)$};

    % Cylinder sides
    \draw[bordism]
        (-1.0,-0.65) -- (-1.0,0.95);

    \draw[bordism]
        (1.0,-0.65) -- (1.0,0.95);

    % Bottom boundary
    \draw[bordism]
        (-1.0,-0.65) arc (180:360:1.0 and 0.28);

    \draw[bordism]
        (-1.0,-0.65) arc (180:0:1.0 and 0.28);

    % Top boundary
    \draw[bordism]
        (-1.0,0.95) arc (180:360:1.0 and 0.28);

    \draw[bordism]
        (-1.0,0.95) arc (180:0:1.0 and 0.28);

    % Boundary labels
    \node[fill=white, inner sep=1.5pt] at (0,-0.6) {$B_1$};
    \node[fill=white, inner sep=1.5pt] at (0,1.0) {$B_2$};

    % N label
    \node at (0,0.15) {$N$};

    % Formula
    \node[align=center] at (0,-1.65)
        {$N\mapsto\mathcal{F}(N):
        \mathcal{F}(B_1)\rightarrow\mathcal{F}(B_2)$};

\end{scope}

\end{tikzpicture}

    \caption{Cartoon depiction of functorial QFT.}
    \label{fig:FuncQFT}
\end{figure}
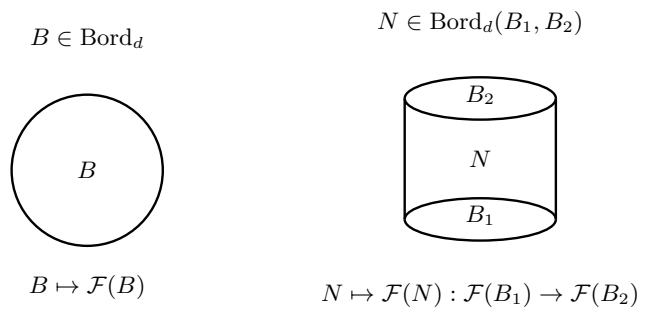

As portrayed in Figure \ref{fig:FuncQFT}, in the standard path integral picture for a $d$-dimensional quantum field theory, we think of Hilbert spaces as being defined on $(d-1)$-dimensional manifolds, and associate $d$-dimensional bulk manifolds with linear transformations between quantum states prepared on their boundaries. The axiomatic approach to QFT formalizes these ideas in terms of concepts from category theory.

This formalization begins with the introduction of the bordism category. The $d$-dimensional bordism category, $\textrm{Bord}_d$, has as its objects $d-1$-dimensional manifolds\footnote{This includes possible operator insertions, background fields, spin structures, orientations and so on. In \cite{McNamara:2026isz} the choice of allowed source data is specified in terms of the choice of a class of manifolds denoted by $\mathcal{X}$. We will suppress this notation for ease of presentation, but we note that the choice of source data has important implications for the analysis. For example, two theories formulated on the same manifolds but endowed with different background field content can be very different!} and as its morphisms $d$-dimensional manifolds whose boundaries are disjoint unions of pairs of $d-1$ manifolds. We will use the notation $B \in \textrm{Bord}_d$ to refer to an object, and $N \in \textrm{Bord}_d(B_1,B_2)$ to refer to a bordism between $B_1, B_2 \in \textrm{Bord}_d$, e.g. $\partial N = B_1 \sqcup B_2$. 

\begin{comment}
Let $\textrm{Bord}_d$ denote the $d$-dimensional bordism category. The objects of $\textrm{Bord}_d$ are $d-1$ dimensional manifolds including possible operator insertions, background fields, spin structures, orientations and so on.\footnote{In \cite{McNamara:2026isz} the choice of allowed source data is specified in terms of the choice of a class of manifolds denoted by $\mathcal{X}$. We will suppress this notation for ease of presentation, but we note that the choice of source data has important implications for the analysis. For example, two theories formulated on the same manifolds but endowed with different background field content can be very different!}  In QFT, these objects will correspond to Hilbert spaces where quantum states are prepared. We will use the notation $B \in \textrm{Bord}_d$ to refer to an object in $\textrm{Bord}_d$. Given two objects $B_1,B_2 \in \textrm{Bord}_d$, a homomorphism $N \in \textrm{Bord}_d(B_1,B_2)$ is a $d$-dimensional manifold with $\partial N = B_1 \sqcup B_2$, compatible with the source data specified on its boundaries. Such a manifold is called a $d$-dimensional bordism between $B_1$ and $B_2$ and in the quantum field theory context will define a linear operator between the Hilbert spaces whose states are specified on $B_1$ and $B_2$. A closed manifold $M \in \textrm{Bord}_d(\emptyset,\emptyset)$ is a special example of a bordism and will be associated with complex numbers through the path integral, e.g. the partition function. 
\end{comment}

Following Atiyah-Segal and Kontsevich-Segal, MW propose the following definition of a $d$-dimensional, unitary quantum field theory \cite[Def. 4.6]{McNamara:2026isz}:

\begin{definition}[$d$-dimensional unitary QFT] \label{def: fQFT}
	A \emph{$d$-dimensional unitary quantum field theory} is a unitary, non-degenerate, symmetric monoidal functor
	\beq
		\mathcal{F}: \textrm{Bord}_d \rightarrow \textrm{Hilb}. 
	\eeq
	This means that $\mathcal{F}$ assigns to each object $B \in \textrm{Bord}_d$ a Hilbert space $\mathcal{F}(B)$, and to each bordism $N \in \textrm{Bord}_d(B_1,B_2)$ a linear operator $\mathcal{F}(N): \mathcal{F}(B_1) \rightarrow \mathcal{F}(B_2)$ satisfying the following properties:
	\begin{enumerate}
		\item \textbf{Functoriality:} $\mathcal{F}(N_1 \circ N_2) = \mathcal{F}(N_1) \circ \mathcal{F}(N_2)$,
		\item \textbf{Non-degeneracy:} Let $C_B(\beta) \in \textrm{Bord}_d(B,B)$ denote the cylinder of size $\beta$ with source and target boundaries $B$, then
		\beq
			\lim_{\beta \rightarrow 0} \mathcal{F}\big(C_B(\beta)\big) = \op{1}_{\mathcal{F}(B)}, 
		\eeq
		\item \textbf{Unitarity:} The category $\textrm{Bord}_d$ comes equipped with an adjoint functor which involves a composition of time reversal and the interchange of the source and target boundaries of a bordism. Denoting this functor by $\dagger$, unitarity implies:
		\beq
			\mathcal{F}(N^{\dagger}) = \mathcal{F}(N)^{\dagger},
		\eeq
		\item \textbf{Symmetric Monoidality} 
		\beq 
			\mathcal{F}(B_1 \sqcup B_2) = \mathcal{F}(B_1) \otimes \mathcal{F}(B_2), \qquad \mathcal{F}(\emptyset) = \mathbb{C}. 
		\eeq
	\end{enumerate}
	
	As a closed manifold, $M$, is a bordism of the empty set to itself and $\mathcal{F}(\emptyset) = \mathbb{C}$, the assignment $\mathcal{F}(M)$ is a scalar which we denote by $Z_{\mathcal{F}}(M)$ and refer to as the partition function.
\end{definition}

So, given a unitary QFT one can always uniquely recover its partition function. On the other hand, one might ask whether it is possible to start from a partition function, satisfying some reasonable properties, and construct a complete unitary QFT. In \cite{McNamara:2026isz}, an affirmative answer is given to this question up an important caveat. The standard construction by cutting open the partition function generally fails to define a unitary QFT due to a breakdown of factorization. The relationship between the QFT obtained by cutting open the partition function and the unitary QFT it is subordinated by can be viewed as a version of Definition \ref{Def Filter}, as we will now describe. 

A $d$-dimensional partition function is a map $Z: \textrm{Bord}_d(\emptyset,\emptyset) \rightarrow \mathbb{C}$. Following the conventions of MW as originally set forward in \cite{Colafranceschi:2023moh}, $Z$ is said to be \emph{finite} if $Z(M) < \infty$ for all $M \in \textrm{Bord}_d(\emptyset,\emptyset)$, \emph{real} if $Z(M^{\dagger}) = \overline{Z(M)}$, \emph{continuous} if $Z(M)$ depends continuously on its domain, and \emph{multiplicative} if $Z(M_1 \sqcup M_2) = Z(M_1) Z(M_2)$. Finally, $Z$ is said to be \emph{reflection positive} if for any $B \in \textrm{Bord}_d$ and any finite collection of $\{M_i\}_{i \in \mathcal{I}} \subset \textrm{Bord}_d(\emptyset,B)$ we have
\beq
	\sum_{i,j \in \mathcal{I}} \bar{c}_i c_j Z(M_i^{\dagger} \circ M_j) \geq 0. 
\eeq

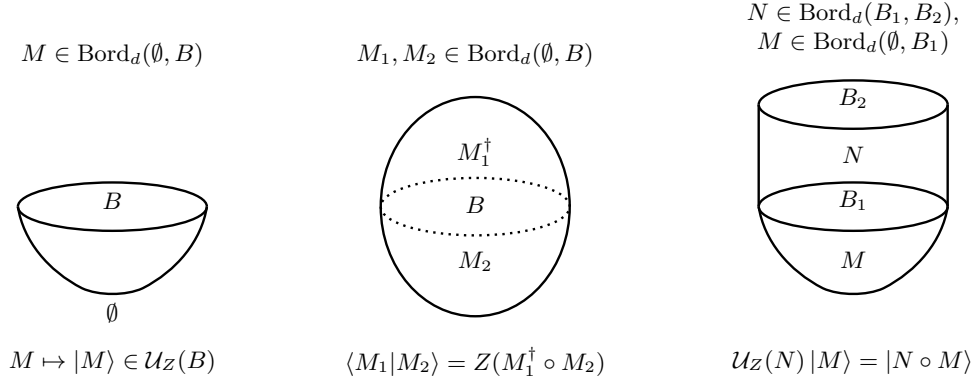
\begin{figure*}[htbp] 
    \centering

\begin{tikzpicture}[
    line width=0.8pt,
    every node/.style={font=\small},
    bordism/.style={draw, line width=0.9pt},
    seam/.style={dotted, line width=0.9pt}
]

% ============================================================
% LEFT
% ============================================================

\begin{scope}[xshift=0cm]

    \node[align=center] at (0,2.0)
        {$M\in\mathrm{Bord}_d(\emptyset,B)$};

    % 3D hemisphere
    % Back half of elliptical boundary
    \draw[bordism]
        (-1.25,0) arc (180:0:1.25 and 0.32);

    % Front half of elliptical boundary
    \draw[bordism]
        (-1.25,0) arc (180:360:1.25 and 0.32);

    % Smooth curved surface
    \draw[bordism]
        (-1.25,0)
        .. controls (-1.18,-0.45) and (-0.85,-0.85) .. (-0.40,-1.08)
        .. controls (-0.18,-1.18) and (0.18,-1.18) .. (0.40,-1.08)
        .. controls (0.85,-0.85) and (1.18,-0.45) .. (1.25,0);

    \node[fill=white, inner sep=1.5pt] at (0,0.08) {$B$};

    \node at (0,-1.40) {$\emptyset$};

    \node[align=center] at (0,-2.05)
        {$M\mapsto\ket{M}\in\mathcal U_Z(B)$};

\end{scope}

% ============================================================
% CENTER
% ============================================================

\begin{scope}[xshift=4.8cm]

    \node[align=center] at (0,2.0)
        {$M_1,M_2\in
        \mathrm{Bord}_d(\emptyset,B)$};

    % Sphere
    \draw[bordism]
        (0,0) ellipse (1.25 and 1.45);

    % Equatorial gluing seam
    \draw[seam]
        (-1.25,0) arc (180:360:1.25 and 0.38);
    \draw[seam]
        (1.25,0) arc (0:180:1.25 and 0.38);

    \node[fill=white, inner sep=1.5pt] at (0,0.03) {$B$};

    \node at (0,0.72) {$M_1^\dagger$};
    \node at (0,-0.72) {$M_2$};

    \node[align=center] at (0,-2.05)
        {$\bra{M_1}\ket{M_2}
        =Z(M_1^\dagger\circ M_2)$};

\end{scope}

% ============================================================
% RIGHT
% ============================================================

\begin{scope}[xshift=9.8cm]

    \node[align=center] at (0,2.35)
        {$\begin{gathered}
        N\in\mathrm{Bord}_d(B_1,B_2),\\[-1mm]
        M\in\mathrm{Bord}_d(\emptyset,B_1)
        \end{gathered}$};

    % Hemisphere
    \draw[bordism]
        (-1.25,0) arc (180:0:1.25 and 0.32);

    \draw[bordism]
        (-1.25,0) arc (180:360:1.25 and 0.32);

    \draw[bordism]
        (-1.25,0)
        .. controls (-1.18,-0.45) and (-0.85,-0.85) .. (-0.40,-1.08)
        .. controls (-0.18,-1.18) and (0.18,-1.18) .. (0.40,-1.08)
        .. controls (0.85,-0.85) and (1.18,-0.45) .. (1.25,0);

    \node[fill=white, inner sep=1.5pt] at (0,0.08) {$B_1$};

    % Cylinder
    \draw[bordism]
        (-1.25,0) -- (-1.25,1.35);

    \draw[bordism]
        (1.25,0) -- (1.25,1.35);

    % Top boundary
    \draw[bordism]
        (-1.25,1.35) arc (180:360:1.25 and 0.32);

    \draw[bordism]
        (-1.25,1.35) arc (180:0:1.25 and 0.32);

    \node[fill=white, inner sep=1.5pt] at (0,1.43) {$B_2$};

    \node at (0,-0.70) {$M$};
    \node at (0,0.68) {$N$};

    \node[align=center] at (0,-2.05)
        {$\mathcal U_Z(N)\ket{M}
        =\ket{N\circ M}$};

\end{scope}

\end{tikzpicture}

    \caption{The universal functor associated wtih  a reflection positive partition function $Z$.}
    \label{fig:universal}
\end{figure*}

Reflection positivity implies that $Z$ can be cut apart to define an inner product and, by extension Hilbert spaces for each $B \in \textrm{Bord}_d$. This is known as the \emph{universal construction} \cite{BLANCHET1995883}, portrayed pictorially in Figure \ref{fig:universal}. Given $Z$, a finite, real, continuous, reflection positive partition function, the Hilbert spaces of the universal construction, $\mathcal{U}_Z(B)$ for each $B \in \textrm{Bord}_d$, are obtained by completing the spaces spanned by states $\ket{M}$ with $M \in \textrm{Bord}_d(\emptyset,B)$ (Left panel of Fig. \ref{fig:universal}) in the pre-inner product $\bra{M_1} \ket{M_2} \equiv Z(M_1^{\dagger} \circ M_2)$ (Center panel of Fig. \ref{fig:universal}). The operators assigned to bordisms, $\mathcal{U}_Z(N)$ for $N \in \textrm{Bord}_d(B_1,B_2)$, are given by the closure of $\mathcal{U}_Z(N) \ket{M} \equiv \ket{N \circ M}$ where $N \circ M$ is the composition of bordisms by gluing along a shared boundary (Right panel of Fig. \ref{fig:universal}). The functor $\mathcal{U}_Z: \textrm{Bord}_d \rightarrow \textrm{Hilb}$ satisfies \cite[Sec. 5.2]{McNamara:2026isz}
\begin{flalign}
	&\lim_{\beta \rightarrow 0} \mathcal{U}_Z(C_B(\beta)) = \op{1}_{\mathcal{U}_Z(B)}, \nonumber \\ 
	&\mathcal{U}_Z(N_1) \circ \mathcal{U}_Z(N_2) = \mathcal{U}_Z(N_1 \circ N_2), \nonumber \\
	&\mathcal{U}_Z(N^{\dagger}) = \mathcal{U}_Z(N)^{\dagger},
\end{flalign}
and is therefore functorial, non-degenerate, and unitary. However, this functor need not, in general, be symmetric monoidal meaning we do not expect factorization
\beq
	\mathcal{U}_{Z}(B_1 \sqcup B_2) \neq \mathcal{U}_Z(B_1) \otimes \mathcal{U}_Z(B_2). 
\eeq

The failure of factorization for the universal construction is quantified by \cite[Thm. 1.1]{McNamara:2026isz}:
\begin{theorem}[McNamara-Wang Reconstruction Theorem] \label{thm non-factorization 1}
	Let $Z$ be a finite, real, continuous, multiplicative, and reflection positive $d$-dimensional partition function with universal construction $\mathcal{U}_Z: \textrm{Bord}_d \rightarrow \textrm{Hilb}$. Then, there exists a unique $d$-dimensional unitary QFT $\mathcal{F}: \textrm{Bord}_d \rightarrow \textrm{Hilb}$ and a compact group $G$ such that
	\begin{flalign} \label{Factorization Restored}
		&Z_{\mathcal{F}} = Z, \nonumber \\ 
		&\mathcal{U}_Z\bigg(B_1 \sqcup ... \sqcup B_n\bigg) = \bigg(\mathcal{F}(B_1) \otimes ... \otimes \mathcal{F}(B_n)\bigg)^G. 
	\end{flalign}
	That is, the partition function of the unitary QFT agrees with $Z$, but the Hilbert space assigned by the universal construction is the $G$-invariant sector of the otherwise factorizing Hilbert space assignment $\mathcal{F}$. 
\end{theorem}

Theorem \ref{thm non-factorization 1} can be reformulated in the language of Definition \ref{Def Filter}. The complete algebras, $\mathscr{A}_{\textrm{ext}}(B) \equiv \mathscr{B}\big(\mathcal{F}(B)\big)$, are the full sets of bounded operators on Hilbert spaces assigned by the associated unitary QFT. Eqn. \eqref{Factorization Restored} tells us that we have an inclusion\footnote{Since we have assumed partition function factorization, this inclusion is in fact an equality when $B = \emptyset$. Otherwise it is strict.}
\beq
	\mathscr{A}_{\textrm{sm}}(B) \subseteq \mathscr{A}_{\textrm{ext}}(B),
\eeq
where the smooth algebras can be identified with the gravitational algebras constructed in \cite{Colafranceschi:2023moh}. In fact, this is a special kind of inclusion since $\mathscr{A}_{\textrm{sm}}$ is the invariant subalgebra of $\mathscr{A}_{\textrm{ext}}$ under an action $\rho: G \rightarrow \text{Aut}(\mathscr{A}_{\textrm{ext}})$. The conditional expectation associated with such an inclusion may be understood as the group averaging map
\beq
	E\big( \mathfrak{X} \big) = \int_{G} d\mu(g) \; \rho_g \big( \mathfrak{X} \big). 
\eeq
Alternatively, the algebra $\mathscr{A}_{\textrm{ext}}$ can be understood as the \emph{crossed product extension} of $\mathscr{A}_{\textrm{sm}}$ with respect to the dual action $\hat{\rho}: L^{\infty}(G) \otimes \mathscr{A}_{\textrm{sm}} \rightarrow \mathscr{A}_{\textrm{sm}}$. This is an instance of a more general result:
\begin{lemma}[Duality between Invariance and Crossed Products] \label{lemma: Inv CP}
	Let $\rho: H \otimes A \rightarrow A$ be the action of a (weak) Hopf algebra $H$ on a $C^*$ algebra $A$, and denote by $A^H \subset A$ the (counital) invariant subalgebra of $A$ under this action. Then, there exists an action $\hat{\rho}: \hat{H} \otimes A^H \rightarrow A^H$ of the dual (weak) Hopf algebra $\hat{H}$ such that $A \simeq A^H \times_{\hat{\rho}} \hat{H}$ \cite{Nill:1998iw}.
\end{lemma}
In our case, the action $\rho$ can be identified with an action $\rho: L^1(G) \otimes \mathscr{A}_{\textrm{ext}} \rightarrow \mathscr{A}_{\textrm{ext}}$, where $L^1(G)$ is the non-Abelian Hopf algebra of the group $G$. Thus, $\mathscr{A}_{\textrm{ext}}$ can be realized as a crossed product of the invariant subalgebra $\mathscr{A}_{\textrm{sm}}$ by the dual action of the dual Hopf algebra, which in this case is $L^{\infty}(G)$. 

\subsection{Application to Quantum Gravity}

To fully match Theorem \ref{thm non-factorization 1} to Definition \ref{Def Filter}, we need to formulate the non-factorization problem for the universal construction in the context of the gravitational path integral. In \cite{Colafranceschi:2023moh}, the authors propose an axiomatic definition of the gravitational path integral for a $D = d+1$-dimensional quantum gravity theory. Inspired by holography, they argue that a $D$-dimensional gravitational path integral is a $d$-dimensional partition function which is finite, real, continuous, multiplicative and reflection positive. It is also possible to relax the assumption of multiplicativity to allow for the possibility that the gravitational path integral does not respect factorization even at the partition function level. We will denote by $\mathcal{U}_{\mathcal{Z}}$ the universal construction of this GPI, and by $\mathscr{A}_{\mathcal{Z}}$ the operator algebra assignment induced by the universal functor. 

Having relaxed multiplicativity, one cannot invoke Theorem \ref{thm non-factorization 1} directly. Instead, we will need to use a more generalized result also discussed in \cite{McNamara:2026isz} and recast in a slightly modified form here:

\begin{conjecture}[Non-Factorization of GPI] \label{thm GPI}
	Let $\mathcal{Z}$ be a finite, real, continous, reflection positive $d$-dimensional partition function representing the GPI of a $D = d+1$-dimensional quantum gravity theory, with $\mathcal{U}_{\mathcal{Z}}$ and $\mathscr{A}_{\mathcal{Z}}$ its associated universal construction and algebra assignment. Then, there exists a $d$-dimensional unitary QFT $\mathcal{F}$ with associated algebraic assignment $\mathscr{A}_{\mathcal{F}}$, and a symmetry algebra $H$ such that
	\beq
		\mathcal{Z} = E(Z_{\mathcal{F}}), \qquad \mathscr{A}_{\mathcal{Z}} = E(\mathscr{A}_{\mathcal{F}}). 
	\eeq
	Here, $E$ is the invariantizing projection (conditional expectation in the algebraic case) associated with the action of $H$.  
\end{conjecture}

Before addressing Conjecture \ref{thm GPI} directly, let us review the case in which $\mathcal{Z}$ is assumed to be multiplicative. In both cases, the construction of $\mathcal{F}$ is achieved by introducing an intermediate category coined by the authors of \cite{McNamara:2026isz} as the baby universe category, $\mathcal{C}_{\textrm{BU}}$. Physically, the baby universe category can be regarded as a `doubling' or `purification' for the universal construction. In the universal construction, states are identified with bordisms from the empty set, e.g. $M \in \textrm{Bord}_d(\emptyset, B)$. In the baby universe category, states are instead identified with general bordisms. As emphasized by MW, from the point of view of the universal construction, such states can be thought of as Einstein-Rosen bridges connecting single boundary Hilbert spaces. The usefulness of the baby universe category, therefore, is that it allows one to `cut open' these Einstein-Rosen bridges, isolating the charged objects which flow between its two boundaries \cite[Sec. 7.1]{McNamara:2026isz}. 

\begin{figure*}[ht]
    \centering

    \begin{tikzpicture}[
        thick,
        line cap=round,
        line join=round
    ]

    % ============================================================
    % LEFT PANEL
    % ============================================================

    \begin{scope}

        \draw[fill=gray!8] (-3,-2) rectangle (3,2);

        % Wilson loop
        \draw[very thick] (0,0) ellipse (2 and 1.15);

        % Bipartition
        \draw[dotted, very thick] (0,-2) -- (0,2);

        \node at (-1.5,2.35) {$\Sigma_A$};
        \node at (1.5,2.35) {$\Sigma_B$};

    \end{scope}

    % ============================================================
    % EQUAL SIGN AND SUM
    % ============================================================

    \node at (3.6,0) {$=$};
    \node at (4.8,0) {$\displaystyle\bigoplus_{\mu}$};

    % ============================================================
    % RIGHT PANEL
    % ============================================================

    \begin{scope}[xshift=9.5cm]

        % Left and right subregions
        \draw[fill=gray!8] (-4,-2) rectangle (-0.8,2);
        \draw[fill=gray!8] (0.8,-2) rectangle (4,2);

        % Left half of Wilson loop: (
        \draw[very thick]
            (-0.8,1.15)
            .. controls (-2.5,1.15) and (-2.5,-1.15) ..
            (-0.8,-1.15);

        % Right half of Wilson loop: )
        \draw[very thick]
            (0.8,1.15)
            .. controls (2.5,1.15) and (2.5,-1.15) ..
            (0.8,-1.15);

        % Charge labels
        \node[right=6pt] at (-0.8,1.15) {$\mu$};
        \node[right=6pt] at (-0.8,-1.15) {$\mu$};

        \node[left=6pt] at (0.8,1.15) {$\bar{\mu}$};
        \node[left=6pt] at (0.8,-1.15) {$\bar{\mu}$};

        % Tensor product
        \node at (0,0) {$\otimes$};

        % Region labels
        \node at (-2.4,2.35) {$\Sigma_A$};
        \node at (2.4,2.35) {$\Sigma_B$};

    \end{scope}

    \end{tikzpicture}

    \caption{In gauge theory, cutting open a Cauchy slice reveals interior charge sectors that otherwise flow unimpeded between disjoint regions.}
    \label{fig:wilsonloop}

\end{figure*}
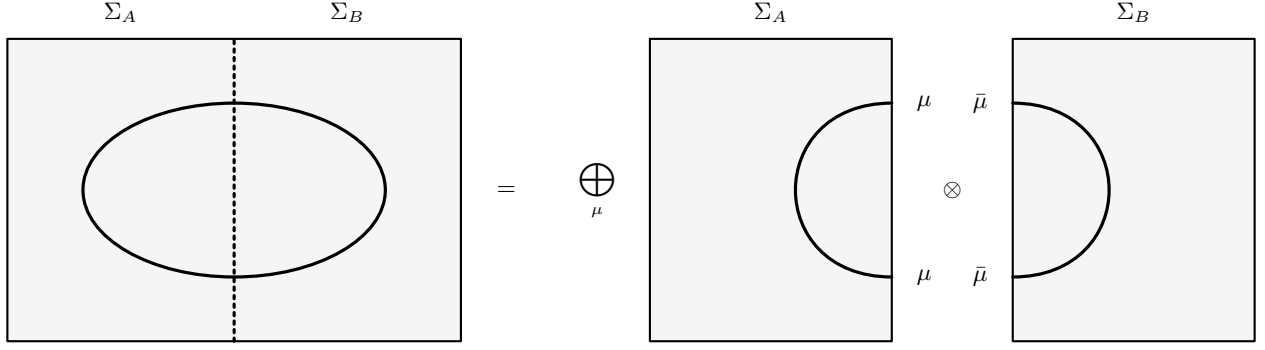

To conceptualize this analysis, it is instructive to recall a similar phenomenology which occurs when one considers the cutting and gluing of subregions in gauge theories \cite{Donnelly:2016auv,Klinger:2023tgi,Ciambelli:2026vxa}. In a complete Cauchy slice, gauge invariant objects -- like Wilson lines -- are extended. When one attempts to identify the physics restricted to a subregion, these extended objects are severed. Extra care must therefore be taken to include new, charged degrees of freedom in the subregion to ensure that it can be consistently glued to complementary regions via a flux conservation condition. The presence of these charges imply a failure of factorization that is structurally analogous to the one described by the baby universe category, see Figure \ref{fig:wilsonloop}. 

\begin{figure*}[htbp]
    \centering

    \begin{tikzpicture}[
    line width=0.8pt,
    every node/.style={font=\small},
    bordism/.style={draw, line width=0.9pt},
    >=stealth
]

% ============================================================
% LEFT PANEL
% ============================================================

\begin{scope}[xshift=0cm]

    \node[align=center] at (0,2.55)
        {$M\in\mathrm{Bord}_d(A,B)$};

    \begin{scope}[yshift=-0.20cm]

        % Cylinder
        \draw[bordism]
            (-1.0,-0.65) -- (-1.0,0.80);
        \draw[bordism]
            (1.0,-0.65) -- (1.0,0.80);

        % Bottom boundary A
        \draw[bordism]
            (-1.0,-0.65) arc (180:360:1.0 and 0.28);
        \draw[bordism]
            (-1.0,-0.65) arc (180:0:1.0 and 0.28);

        % Top boundary B
        \draw[bordism]
            (-1.0,0.80) arc (180:360:1.0 and 0.28);
        \draw[bordism]
            (-1.0,0.80) arc (180:0:1.0 and 0.28);

        \node[fill=white, inner sep=1.5pt] at (0,-0.6) {$A$};
        \node[fill=white, inner sep=1.5pt] at (0,0.85) {$B$};

        \node at (0,0.10) {$M$};

    \end{scope}

    \node[align=center] at (0,-1.95)
        {$\mathcal U_{\mathcal{Z}}(M):
        \mathcal U_{\mathcal{Z}}(A)\rightarrow\mathcal U_{\mathcal{Z}}(B)$};

\end{scope}

% ============================================================
% ARROW
% ============================================================

\begin{scope}[xshift=2.75cm]
    \node at (0,0.10) {$\mapsto$};
\end{scope}

% ============================================================
% CENTER PANEL
% ============================================================

\begin{scope}[xshift=5.4cm]

    %\node[align=center] at (0,2.55)
        %{$M$};

    \begin{scope}[yshift=-0.20cm]

        % Outer wall
        \draw[bordism]
            (-1.15,0.65)
            -- (-1.15,-0.35)
            .. controls (-1.15,-1.15) and (-0.55,-1.45) .. (0,-1.45)
            .. controls (0.55,-1.45) and (1.15,-1.15) .. (1.15,-0.35)
            -- (1.15,0.65);

        % Inner wall
        \draw[bordism]
            (-0.55,0.65)
            -- (-0.55,-0.30)
            .. controls (-0.55,-0.70) and (-0.25,-0.85) .. (0,-0.85)
            .. controls (0.25,-0.85) and (0.55,-0.70) .. (0.55,-0.30)
            -- (0.55,0.65);

        % Left boundary
        \draw[bordism]
            (-1.15,0.65) arc (180:360:0.30 and 0.20);
        \draw[bordism]
            (-1.15,0.65) arc (180:0:0.30 and 0.20);

        % Right boundary
        \draw[bordism]
            (0.55,0.65) arc (180:360:0.30 and 0.20);
        \draw[bordism]
            (0.55,0.65) arc (180:0:0.30 and 0.20);

        \node[fill=white, inner sep=2pt] at (-1.45,0.75) {$B$};
        \node[fill=white, inner sep=2pt] at (1.45,0.75) {$\bar A$};

        \node at (0,-0.40) {$M$};

    \end{scope}

    \node[align=center] at (0,-2.15)
        {$\ket{M}\in
        \mathcal U_{\mathcal{Z}}(B\sqcup\bar A)$};

\end{scope}

% ============================================================
% RIGHT PANEL
% ============================================================

\begin{scope}[xshift=11.0cm]

    \node[align=center] at (0,2.55)
        {$N\in\mathrm{Bord}_d(B,B')$};

    \begin{scope}[yshift=-0.20cm]

        % Horseshoe
        \draw[bordism]
            (-1.15,0.65)
            -- (-1.15,-0.35)
            .. controls (-1.15,-1.15) and (-0.55,-1.45) .. (0,-1.45)
            .. controls (0.55,-1.45) and (1.15,-1.15) .. (1.15,-0.35)
            -- (1.15,0.65);

        \draw[bordism]
            (-0.55,0.65)
            -- (-0.55,-0.30)
            .. controls (-0.55,-0.70) and (-0.25,-0.85) .. (0,-0.85)
            .. controls (0.25,-0.85) and (0.55,-0.70) .. (0.55,-0.30)
            -- (0.55,0.65);

        % Right boundary
        \draw[bordism]
            (0.55,0.65) arc (180:360:0.30 and 0.20);
        \draw[bordism]
            (0.55,0.65) arc (180:0:0.30 and 0.20);

        % N
        \draw[bordism]
            (-1.15,0.65) -- (-1.15,1.65);
        \draw[bordism]
            (-0.55,0.65) -- (-0.55,1.65);

        % B'
        \draw[bordism]
            (-1.15,1.65) arc (180:360:0.30 and 0.20);
        \draw[bordism]
            (-1.15,1.65) arc (180:0:0.30 and 0.20);

        % Gluing seam B
        \draw[dotted, line width=1pt]
            (-1.15,0.65) arc (180:360:0.30 and 0.20);
        \draw[dotted, line width=1pt]
            (-1.15,0.65) arc (180:0:0.30 and 0.20);

        % Labels
        \node[fill=white, inner sep=2pt] at (-1.45,0.75) {$B$};
        \node[fill=white, inner sep=2pt] at (-1.45,1.65) {$B'$};
        \node[fill=white, inner sep=2pt] at (1.45,0.75) {$\bar A$};

        \node at (-0.85,1.15) {$N$};
        \node at (0,-0.40) {$M$};

    \end{scope}

    \node[align=center] at (0,-2.15)
        {$\mathcal U_{\mathcal{Z}}^{(L)}(N)\ket{M}
        =\ket{N\circ_L M}$};

\end{scope}

\end{tikzpicture}

    \caption{In the baby universe category, bordisms $M \in \textrm{Bord}_d(A,B)$ are mapped to states in the `doubled' Hilbert space $\mathcal{U}_{\mathcal{Z}}(B \sqcup \bar{A})$. They can then be acted upon by composition of bordisms along either the left or right open boundary. See also \cite[Fig. 13]{McNamara:2026isz}}
    \label{fig:BUCat}
\end{figure*}
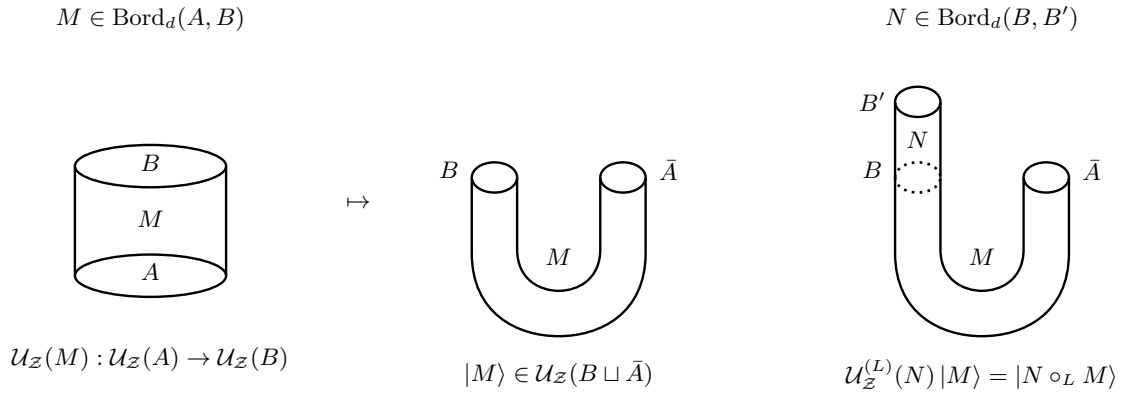

The construction of the baby universe category begins by studying the two-boundary Hilbert spaces $\mathcal{U}_{\mathcal{Z}}(B_1  \sqcup \overline{B_2})$. The states in $\mathcal{U}_{\mathcal{Z}}(B_1 \sqcup \overline{B_2})$ may be identified with bordisms $M \in \textrm{Bord}_d(B_2,B_1)$ whereupon the lower boundary is swung around to prepare a state with conjugate orientation \cite[Sec. 7.1]{McNamara:2026isz}, see Figure \ref{fig:BUCat}.\footnote{This is perfectly analogous to the standard observation that an operator $\mathcal{O}: \mathscr{H}_1 \rightarrow \mathscr{H}_2$ can equivalently be regarded as a state $\ket{\mathcal{O}} \in \mathscr{H}_2 \otimes \mathscr{H}_1^*$.} It is in this sense that one may think of $\mathcal{U}_{\mathcal{Z}}(B_1 \sqcup \overline{B_2})$ as describing the Hilbert space of Einstein-Rosen bridges between $B_1$ and $B_2$. The universal construction then furnishes a `representation' of $\textrm{Bord}_d$ acting on these Hilbert spaces:
\beq \label{Rep}
	\mathcal{U}^{(L)}_{\mathcal{Z}}(N) \ket{M} \equiv \ket{N \circ_L M}. 
\eeq
Here, $\ket{M} \in \mathcal{U}_{\mathcal{Z}}(B_1 \sqcup \overline{B_2})$ is identified with the bordism $M \in \textrm{Bord}_d(B_2,B_1)$, $N \in \textrm{Bord}_d(B_1, A)$, and the notation $N \circ_L M$ indicates a composition of bordisms along the \emph{left}\footnote{Analogously, we can define a representation $\mathcal{U}^{(R)}_{\mathcal{Z}}$ in which bordisms act on the right boundary.} open boundary of $\ket{M}$. This means that $N \circ_L M \in \textrm{Bord}_d(B_2,A)$ and thus $\ket{N \circ_L M} \in \mathcal{U}_{\mathcal{Z}}(A \sqcup \overline{B_2})$. 

In addition to the representation \eqref{Rep}, the assignment $(B_1,B_2) \mapsto \mathcal{U}_{\mathcal{Z}}(B_1 \sqcup \overline{B_2})$ defines a Hilbert space valued inner product on the bordism cateogry \cite[Eqn. 7.29]{McNamara:2026isz}
\begin{flalign} \label{Inner}
	&\langle \langle \cdot, \cdot \rangle \rangle: \textrm{Bord}_d^{\times 2} \rightarrow \textrm{Hilb}, \nonumber \\
	&\langle\langle B_1, B_2 \rangle \rangle \equiv \mathcal{U}_{\mathcal{Z}}(B_1 \sqcup \overline{B_2}). 
\end{flalign}
The baby universe category, $\mathcal{C}_{\textrm{BU}}$, is formally obtained by (i) taking a categorical analog of weak closure \cite{GhezLimaRoberts1985,henriques2024completewcategories,bunke2025additiveccategoriesktheory} with respect to to the representation \eqref{Rep} and (ii) taking a categorical analog of Cauchy completion \cite{Binder:2019zqc,henriques2024completewcategories} with respect to the inner product \eqref{Inner}. This two step process can be regarded as a generalization of the GNS construction for $C^*$ algebras to $C^*$ categories \cite[Sec. 7]{McNamara:2026isz}.

The objects of $\mathcal{C}_{\textrm{BU}}$ can be understood as formal linear combinations \cite[Eqn. 7.37]{McNamara:2026isz}
\beq
	\Psi = \bigoplus_{i} \mathscr{H}_i \otimes B_i, \qquad \mathscr{H}_i \in \textrm{Hilb}, \; B_i \in \textrm{Bord}_d. 
\eeq
Let us denote the irreducible objects of $\mathcal{C}_{\textrm{BU}}$ by $\mathscr{H}_{\mu}$, which MW identify with the $\mu$-sectors of \cite{Colafranceschi:2023moh}. By \cite[Sec. 7.4.1]{McNamara:2026isz}, the $\mu$-sectors form a categorical orthonormal basis in the sense that \cite[Eqn. 7.53, 7.54]{McNamara:2026isz}
\beq
	\langle \langle \mathscr{H}_{\mu}, \mathscr{H}_{\nu} \rangle \rangle = \delta_{\mu \nu} \mathbb{C}, \qquad \textrm{id}_{\mathcal{C}_{\textrm{BU}}} = \bigoplus_{\mu} \mathscr{H}_{\mu} \otimes \mathscr{H}_{\mu}^*. 
\eeq
This allows us to decompose the Einstein-Rosen bridge as \cite{Colafranceschi:2023moh,Marolf:2024adj}
\begin{flalign} \label{Failure of Factorization from CBU}
	\mathcal{U}_{\mathcal{Z}}(B_1 \sqcup \overline{B_2}) &= \langle \langle B_1, B_2 \rangle \rangle \nonumber \\
	&= \bigoplus_{\mu} \langle \langle B_1, \mathscr{H}_{\mu} \rangle \rangle \langle \langle \mathscr{H}_{\mu}, B_2 \rangle \rangle \nonumber \\
	&= \bigoplus_{\mu} \mathscr{H}_{B_1}^{\mu} \otimes \mathscr{H}_{B_2}^{\mu}.
\end{flalign}
In this sense, we can think of $\mathscr{H}_{\mu}$ as half-ER bridges \cite{Saad:2021rcu,Colafranceschi:2023moh,McNamara:2026isz}, analogous to the half Wilson loops in Figure \ref{fig:wilsonloop}. This is particularly well illustrated in \cite[Fig. 15]{McNamara:2026isz}.

Given the baby universe category $\mathcal{C}_{\textrm{BU}}$, the construction of the unitary QFT follows simply from two observations. First, there is a canonical unitary, symmetric monoidal functor $\mathcal{U}_{\mathcal{Z}}: \textrm{Bord}_d \rightarrow \mathcal{C}_{\textrm{BU}}$. Here, by a slight abuse of notation, we have referred to this functor by $\mathcal{U}_{\mathcal{Z}}$ to emphasize its relation to the universal construction. More rigorously, it arises from the sequence of assignments described above. Second, by \cite[Section 8-9]{McNamara:2026isz} the baby universe category is rigidly generated \cite[Def. 9.1]{McNamara:2026isz}, such that the Doplicher-Roberts theorem \cite{Doplicher:1971wk,Doplicher:1973at,Doplicher:1989afv} provides a fiber functor $\Phi: \mathcal{C}_{\textrm{BU}} \rightarrow \textrm{Hilb}$. The existence of the fiber functor $\Phi$ implies that $\mathcal{C}_{\textrm{BU}}$ can be regarded as equivalent to the representation category of a compact group $G$. The $\mu$-sectors of $\mathcal{C}_{\textrm{BU}}$ are therefore identified with the irreducible representations of this group. The composition of $\mathcal{U}_{\mathcal{Z}}$ and $\Phi$ yields a unitary quantum field theory $\mathcal{F}: \Phi \circ \mathcal{U}_{\mathcal{Z}}: \textrm{Bord}_d \rightarrow \textrm{Hilb}$, whose partition function, $Z_{\mathcal{F}}$, is equivalent to $Z$ \cite[Thm. 1.1(c)]{McNamara:2026isz}. The act of post-composing with the fiber functor $\Phi$ can be regarded as appending to the universal construction those states charged under $G$ whose omission led to the failure of factorization \eqref{Failure of Factorization from CBU}. 

This state of affairs can equivalently be interpreted as constructing the net of inclusions $\mathscr{A}_{\textrm{sm}}(B) \subseteq \mathscr{A}_{\textrm{ext}}(B)$ with associated conditional expectation $E$. From this point of view, the baby universe category is analogous to the bimodule category of the inclusion \cite{nikshych2000galoiscorrespondenceii1factors}. The fact that $\mathscr{A}_{\textrm{sm}}$ is realized as the $G$-invariant algebra of $\mathscr{A}_{\textrm{ext}}$ implies that this bimodule category is equivalent to the representation category of $G$. The extended algebra can be understood as the algebraic union of the smooth algebra with a set of charged intertwiners carrying the fusion structure of the bimodule category of the inclusion. In this case, these intertwiners encode the fusion algebra of $\textrm{Rep}(G)$, which extends to the commutative group algebra $L^{\infty}(G)$, thereby recovering our observation that $\mathscr{A}_{\textrm{ext}} \simeq \mathscr{A}_{\textrm{sm}} \rtimes L^{\infty}(G)$. In this sense, the baby universe category can be regarded as a device for determining the charged intertwiners naturally associated with the failure of factorization for the smooth gravitational theory. 

So, what changes when we relax multiplicativity? The authors of \cite{McNamara:2026isz} address this case, too. The failure of multiplicativity implies the existence of a family of multiplicative partition functions $\{\mathcal{Z}_{\alpha}\}$ and a probability measure $d\mu(\alpha)$ such that
\beq \label{alpha sectors}
	\mathcal{Z}(M_1 \sqcup ... \sqcup M_n) = \int d\mu(\alpha) \; \mathcal{Z}_{\alpha}(M_1) \cdot ... \cdot \mathcal{Z}_{\alpha}(M_n).
\eeq
Of course, $\alpha$ are the famous $\alpha$-sectors of \cite{Giddings:1988wv,Marolf:2020xie}. Given this observation, we can run the above analysis for each factorizing $\alpha$-sector to obtain a family of unitary quantum field theories $\{\mathcal{F}_{\alpha}\}$ and compact groups $\{G_{\alpha}\}$ such that the Hilbert spaces assigned by the universal construction are of the form\footnote{We should emphasize that the status of these results for the case in which there is a diffuse measure on the space of $\alpha$-sectors is conjectural, as discussed in \cite[Section 10.3]{McNamara:2026isz} and reiterated in Conjecture \ref{Lemma} below. It is proven rigorously in the case that the measure is discrete.} \cite[Eqn. 10.42]{McNamara:2026isz}
\begin{flalign} \label{Non Mult Fact}
	\mathcal{U}_{\mathcal{Z}} & \bigg(B_1 \sqcup ... \sqcup B_n\bigg)  \nonumber \\
	&= \int^{\oplus} d\mu(\alpha) \; \bigg(\mathcal{F}_{\alpha}(B_1) \otimes ... \otimes \mathcal{F}_{\alpha}(B_n)\bigg)^{G_{\alpha}}. 
\end{flalign}

More technically, when $\mathcal{Z}$ fails to be multiplicative, the baby universe category -- obtained by the same construction outlined above -- is promoted from a rigidly generated $W^*$-tensor category to a rigidly generated $W^*$-\emph{multi}tensor category \cite[Def. 9.1]{McNamara:2026isz}. Heuristically, this means that
\beq \label{Multi BU}
	\mathcal{C}_{\textrm{BU}} = \int^{\oplus} d\mu(\alpha) \; \mathcal{C}_{\textrm{BU}}^{\alpha},
\eeq	 
where $\mathcal{C}_{\textrm{BU}}^{\alpha}$ is the baby universe category induced from the universal construction of the $\alpha^{th}$ factorizing partition function $\mathcal{Z}_{\alpha}$. This implies that the unit object in the category $\mathcal{C}_{\textrm{BU}}$ is not irreducible. As we shall see in a moment, this observation implies that even the empty set (e.g. closed universe) can be cut open to reveal non-trivial charge sectors. 

Nevertheless, the universal construction continues to define a unitary, symmetric monoidal functor $\mathcal{U}_{\mathcal{Z}}: \textrm{Bord}_d \rightarrow \mathcal{C}_{\textrm{BU}}$. Following the approach  applied in the multiplicative case, we would next like to argue that we can complete this functor to a unitary QFT by post-composing with a fiber functor $\Phi: \mathcal{C}_{\textrm{BU}} \rightarrow \textrm{Hilb}$. However, there is an immediate obstruction: the unit object in $\mathcal{C}_{\textrm{BU}}$ is no longer irreducible, while the unit object in $\textrm{Hilb}$ -- $\mathbb{C}$ -- is. This implies that there can be no single fiber functor which sees the entire baby universe category \cite[Sec. 9.1]{McNamara:2026isz}.

There is important physics hiding in this observation. It tells us that it is not possible to embed a GPI into a single functor taking values in $\textrm{Hilb}$. However, it does not necessarily prohibit the existence of a single unitary QFT into which the GPI can be embedded, provided one is willing to generalize slightly what one means by a quantum field theory. In particular, following MW, we may consider a QFT to be a map $\mathcal{F}: \textrm{Bord}_d \rightarrow \mathcal{D}$, where for the moment $\mathcal{D}$ is a general category. From this point of view, the functor $\mathcal{U}_{\mathcal{Z}}: \textrm{Bord}_d \rightarrow \mathcal{C}_{\textrm{BU}}$ defines a $\mathcal{C}_{\textrm{BU}}$-valued QFT. In this sense, $\mathcal{C}_{\textrm{BU}}$ can be regarded as the canonical category into which the original theory can be completed to a QFT which is no longer valued in $\textrm{Hilb}$. Unraveling the physical consequences of this observation is the task we turn to next.

\section{Reducible Quantum Field Theories and Weak Hopf Algebraic Symmetries} \label{sec: WHA}

The preceding discussion naturally emphasizes the \emph{ensemble} nature of the non-factorizing gravitational path integral e.g. in the spirit of Marolf-Maxfield. What we would now like to argue is that it is also possible to arrange the degrees of freedom associated with the $\alpha$-sectors and the $\mu$-sectors into a \emph{single} theory with a \emph{single} factorizing partition function and factorizing Hilbert spaces. 

\subsection{A single theory to fit all $\alpha$ sectors} \label{sec: PFact}

To begin, let us develop a bit more what it would mean to have a QFT with values in the category $\mathcal{D}$. As we would like to retain the standard quantum mechanical interpretation, we will demand that $\mathcal{D}$ still has as its objects Hilbert spaces and as its homomorphisms linear maps between Hilbert spaces. Nevertheless, the monoidal structure in the category can be amended e.g. such that the unit object is no longer irreducible. The axioms in Definition \ref{def: fQFT} remain unchanged with the exception of symmetric monoidality which becomes:
\beq \label{D-functoriality}
	\mathcal{F}(B_1 \sqcup B_2) = \mathcal{F}(B_1) \otimes_{\mathcal{D}} \mathcal{F}(B_2), \;\; \mathcal{F}(\emptyset) = \mathscr{H}_{\textrm{BU}}. 
\eeq
That is, we continue to demand factorization (crucially with respect to the tensor product operation native to $\mathcal{D}$) but allow the empty set to be mapped to a generally non-trivial Hilbert space we have identified with the space of baby universe states.  A functor $\mathcal{F}: \textrm{Bord}_d \rightarrow \mathcal{D}$ satisfying the $\mathcal{D}$-adjusted axioms will be called a $d$-dimensional \emph{reducible}\footnote{The term \emph{relative}, as formulated in \cite{Freed:2012bs}, may also be appropriate here. We thank Jake McNamara for bringing this to our attention.} unitary QFT. 

For a standard, irreducible, unitary quantum field theory, we used the fact that $\mathcal{F}(\emptyset) = \mathbb{C}$ to argue that, for any $M \in \textrm{Bord}_d(\emptyset,\emptyset)$, $\mathcal{F}(M) \in \mathscr{B}(\mathbb{C}) \simeq \mathbb{C}$ could be identified simply with a complex number we interpreted as the partition function. In this more general case, we see that a closed manifold is mapped to an operator $\mathcal{F}(M) \in \mathscr{B}(\mathscr{H}_{\textrm{BU}})$. Since $\mathscr{H}_{\textrm{BU}}$ is the identity object in $\mathcal{D}$, we have $\mathscr{H}_{\textrm{BU}} \otimes_{\mathcal{D}} \mathscr{H}_{\textrm{BU}} \simeq \mathscr{H}_{\textrm{BU}}$, and in particular $\mathcal{F}$ assigns to the disjoint union of closed manifolds the product of operators on $\mathscr{H}_{\textrm{BU}}$: 
\beq \label{Op Part}
	\mathcal{F}(M_1 \sqcup M_2) = \mathcal{F}(M_1) \mathcal{F}(M_2). 
\eeq

On one hand, the fact that the identity object is no longer $\mathbb{C}$ can be interpreted as a `change of scalars'. In other words, anywhere there was complex numbers we replace them by elements in the baby universe Hilbert space. The partition function, which was an element of $\mathscr{B}(\mathbb{C}) \simeq \mathbb{C}$, would then become an operator on $\mathscr{H}_{\textrm{BU}}$ -- as in \eqref{Op Part}. More generally, states associated with boundaries are `tensored' with $\mathscr{H}_{\textrm{BU}}$ and operators associated with bordisms are `tensored' with $\mathscr{B}(\mathscr{H}_{\textrm{BU}})$. In some sense, this point of view is quite consistent with the ensemble perspective, albeit with one important caveat. The `scalars' $\mathscr{H}_{\textrm{BU}}$ can roughly be thought of as encoding the partition functions, states, and operators for the theory in every $\alpha$ sector all at once rather than only after averaging. 

This latter observation suggests that the extended theory is somewhat more than just an ensemble average. All of the $\alpha$-sectors coexist within a single system. If we think of the $\alpha$-sectors as weakly interacting subsystems in a single theory, the partition function we would assign to them would be the product of their individual partition functions. More invariantly, we can  associate a single complex value partition function to this theory by taking the determinant:
\beq \label{Gen Partition Function}
	Z_{\mathcal{F}}(M) = \det(\mathcal{F}(M)). 
\eeq
Since the determinant of a product of operators is equal to the product of determinants we find that 
\beq
	Z_{\mathcal{F}}(M_1 \sqcup M_2) = Z_{\mathcal{F}}(M_1) Z_{\mathcal{F}}(M_2), 
\eeq
and so factorization is preserved.  

Given this perspective, the existence of a single unitary QFT extending $\mathcal{U}_{\mathcal{Z}}$ again comes down simply to the question of whether $\mathcal{C}_{\textrm{BU}}$, as a rigidly generated symmetric $W^*$-multitensor category, admits a functor into a category $\mathcal{D}$ that extends $\textrm{Hilb}$ in the manner described above. Given such a functor, $\Phi: \mathcal{C}_{\textrm{BU}} \rightarrow \mathcal{D}$, we automatically obtain the desired unitary quantum field theory as $\mathcal{F} \equiv \Phi \circ \mathcal{U}_{\mathcal{Z}}$. A large class of $\mathcal{D}$ which fit the bill can be obtained from the representation categories of symmetry algebras. Conjecture \ref{thm GPI} therefore collapses to the following Conjecture:

\begin{conjecture}[Reconstruction for Multitensor Categories] \label{Lemma}
	Given a rigidly generated symmetric $W^*$-multitensor category $\mathcal{C}$, there exists a symmetry algebra $H$ such that $\mathcal{C} = \text{Rep}(H)$. 
\end{conjecture}

In general, the validity of this conjecture is the subject of ongoing analysis for rather technical reasons as anticipated in \cite[Section 10.3]{McNamara:2026isz}.\footnote{We note, if the assumption of symmetry for the category is relaxed the problem of reconstruction for rigidly generated $W^*$ multitensor categories could be related to an algebraic formalism for continuous non-invertible symmetries \cite{AliAhmad:2025bnd}.} However, in the case that $\mathcal{C}$ is a multi\emph{fusion} rather than a multi\emph{tensor} category, it is known to be true \cite{Ostrik:2001xnt,Etingof:2002vpd,Bai:2025zze,AliAhmad:2025bnd}.\footnote{We note that the uniqueness of this reconstruction is up to categorical, or Morita equivalence. The Morita equivalence class of weak Hopf algebras reconstructing $\mathcal{C}$ is governed by module category structure of $\mathcal{C}$. This has interesting physical implications, discussed e.g. in \cite{AliAhmad:2025bnd}. We plan to explore the role of Morita equivalence for the algebraic filter in future work.} In that case the symmetry algebra $H$ is a \emph{weak Hopf algebra}. Thus, when $\mathcal{C}_{\textrm{BU}}$ is a multifusion category, we conclude that there exists a unique (up to Morita equivalence) \emph{single} unitary quantum field theory extending $\mathcal{Z}$ which takes values in the category $\textrm{Rep}(H)$ for $H$ a weak Hopf algebra reconstructed from $\mathcal{C}_{\textrm{BU}}$. In fact, due to the symmetric structure of the baby universe category, this weak Hopf algebra can be realized as the algebra of a classical \emph{groupoid}, $\mathcal{G}$ \cite{McNamara:2026isz}. 

\subsection{Weak Hopf Algebras}

Given the affirmative form of Conjecture \ref{Lemma} in the multifusion case, we see that, including the effects of partition function non-factorization, the unitary QFT $\mathcal{F}$ is obtained by extending the universal construction by charged states associated with a groupoid rather than a group. More broadly, this implies that the unitary QFT which completes the GPI is \emph{reducible}, taking values in a category whose unit object is non-trivial. 

In this section, we will briefly explore the structure of reducible QFTs through their relation to general weak Hopf algebras. Weak Hopf algebras are sometimes referred to as \emph{quantum groupoids} in the sense that they `quantize' the structural properties of groupoids. In particular, weak Hopf algebras provide a general language for understanding the consequences of having a reducible unit object in the theory, e.g. a non-trivial baby universe Hilbert space. The gravitational construction of \cite{McNamara:2026isz} is an example of such a theory in which the weak Hopf algebra is \emph{classical}. As described in \cite{AliAhmad:2025bnd}, we expect a genuinely weak structure to arise for systems with exotic vacuum structures and non-invertible symmetries. 

Formally, a weak Hopf algebra is a unital, $C^*$ algebra $H$ together with a coproduct $\Gamma: H \rightarrow H \otimes H$, a counit $\epsilon: H \rightarrow \mathbb{C}$ and an antipode $S: H \rightarrow H$ satisfying a set of compatibility conditions \cite{Rehren:1996ku,nikshych2002finite,
nikshych1999dualitytheoremquantumgroupoids,
BOHM1999385,bohm1999weakhopfalgebrasii,Nill:1998iw,AliAhmad:2025bnd}. If the coproduct is a unital map, $H$ is called simply a \emph{Hopf algebra}. In a weak Hopf algebra, the nonunitality of the coproduct implies the existence of a pair of projections
\beq
	\epsilon_t(h) \equiv h_{(1)} S(h_{(2)}), \qquad \epsilon_s(h) \equiv S(h_{(1)}) h_{(2)}
\eeq
called the counital target and source maps.\footnote{Here we have employed Sweedler notation
\beq
	\Gamma(h) = h_{(1)} \otimes h_{(2)},
\eeq
with an implied summation over terms.} The images of these maps are the target and source subalgebras, $H_t \equiv \epsilon_t(H)$ and $H_s \equiv \epsilon_s(H)$. For a Hopf algebra, $\epsilon_s$ and $\epsilon_t$ are each equivalent to the counit $\epsilon$, and thus the target and source subalgebra trivialize as $H_t = H_s = \mathbb{C}$. 

The existence of non-trivial source and target counital subalgebras is therefore the defining feature of a weak Hopf algebra \cite{Etingof:2002vpd}. We can also recognize this difference from the categorical point of view. The category $\textrm{Rep}(H)$ has as its objects Hilbert space representations of $H$, and as its homs intertwining operators between representations. This category can be endowed with a monoidal structure which differs between whether $H$ is a genuine Hopf algebra or merely a weak Hopf algebra. When $H$ is Hopf, the tensor product can be taken to be the ordinary tensor product of representations, and the unit object is the trivial representation, which is irreducible. By contrast, when $H$ is weak Hopf, the monoidal functor is relative tensor product over the target subalgebra $H_t$. The GNS Hilbert space of $H_t$ also plays the role of the unit object, which is reducible as a representation. Given our definition, we therefore recognize that a reducible QFT generically will take values in the representation category of a weak Hopf algebra $H$, with its (analog of the) baby universe Hilbert space equivalent to the target subalgebra $H_t$. 

To make the difference between Hopf algebras and \emph{weak} Hopf algebras concrete, it is instructive to consider the difference between algebras associated with groups and algebras associated with groupoids. Given a compact group $G$ we can define two standard examples of weak Hopf algebras which are dual\footnote{Given a weak Hopf algebra, $H$, there exists a natural dual weak Hopf algebra $\hat{H}$ identified with the vector space dual to $H$ e.g. $\hat{H} = \{\alpha: H \rightarrow \mathbb{C} \; | \; \textrm{linear}\}$. The structure maps of $\hat{H}$ are induced from those of $H$ by duality
\begin{flalign}
	&(\alpha \beta, h) \equiv (\alpha \otimes \beta, \Gamma(h)), \qquad (\hat{\op{1}}, h) \equiv \epsilon(h), \nonumber \\
	&(\hat{\Gamma}(\alpha), h \otimes g) \equiv (\alpha,hg), \qquad \hat{\epsilon}(\alpha) \equiv (\alpha,\op{1}), \nonumber \\
	&(\hat{S}(\alpha),h) \equiv (\alpha,S(h)), \qquad (\alpha^*, h) \equiv (\alpha,S(h)^*). 
\end{flalign}
Here, $(,): \hat{H} \otimes H \rightarrow \mathbb{C}$ is the dual pairing.} to each other. The first is the commutative group algebra $L^{\infty}(G)$ with structure maps
\begin{flalign} \label{AHA G}
	&\bigg(f_1 \cdot f_2\bigg)(g) \equiv f_1(g) f_2(g), \nonumber \\ 
	&\Gamma(f)(g_1,g_2) \equiv f(g_1 g_2), \nonumber \\ 
	&\epsilon(f) = f(e), \nonumber \\ 
	&S(f)(g) \equiv f(g^{-1}). 
\end{flalign}
Likewise, there is the noncommutative group algebra $L^1(G)$ with structure maps
\begin{flalign} \label{NAHA G}
	&\bigg(f_1 \star f_2\bigg)(g) = \int_{G} d\mu(h) f_1(h) f_2(h^{-1} g), \nonumber \\ 
	&\Gamma(f)(g_1,g_2) = f(g_1 g_2), \nonumber \\ 
	&\epsilon(f) = \int_{G} d\mu(g) f(g), \nonumber \\ 
	&S(f)(g) = f(g^{-1}). 
\end{flalign}
It is straightforward to see that the coproducts for these algebras are unital, and thus $L^{\infty}(G)$ and $L^1(G)$ are ordinary, rather than \emph{weak} Hopf algebras. 

By contrast, the standard examples of genuine weak Hopf algebras are the commutative and noncommutative algebras associated with a group\emph{oid}. Recall that a groupoid is a small category, $\mathcal{G}$, whose homomorphisms are all invertible, and thus can be thought of as the many-object generalization of a group. The set of objects in $\mathcal{G}$, which we denote by $\mathcal{G}_0$, is referred to as the base of the groupoid and signifies the possible targets and sources of groupoid transformations. One can form commutative and noncommutative generalizations of \eqref{AHA G} and \eqref{NAHA G} for a groupoid based on $L^{\infty}(\mathcal{G})$ and $L^1(\mathcal{G})$, respectively \cite{nikshych2002finite}. The resulting objects are genuine \emph{weak} Hopf algebras whose target and source counital algebras coincide with $L^{\infty}(\mathcal{G}_0)$. 

One can always obtain a groupoid by forming a bundle of groups e.g.
\beq
	\mathcal{G} = \int^{\oplus}_{X} d\mu(\alpha) \; G_{\alpha},
\eeq
where $(X,d\mu)$ is a measure space. Of course, this is precisely the symmetry structure expected from the baby universe category after relaxing multiplicativity \cite{McNamara:2026isz}. Thus, the target and source counital algebras naturally encode the information of the $\alpha$-sectors in the generalized reconstruction of Theorem \ref{thm GPI}. Specifically, the baby-universe Hilbert space in this case is given by $\mathscr{H}_{\textrm{BU}} = L^2(\mathcal{G}_0) = L^2(X,d\mu)$ \cite[Eqn. 9.28]{McNamara:2026isz}. 

Reference \cite{AliAhmad:2025bnd} reviews the theory of weak Hopf covariant systems which describes the action of weak Hopf algebras. Given a weak Hopf action $\rho: H \otimes S \rightarrow S$, we can always define a projection to the `invariant' subset $S^H$ by applying the analog of averaging over a group action with respect to its invariant Haar measure. The object that replaces the Haar measure is the Haar integral element $\Lambda \in H$. For a finite $C^*$ weak Hopf algebra the Haar integral can always be written in the form \cite{BOHM1999385}
\beq
	\Lambda = \sum_{U \in \textrm{IrrRep}(H_t)} \sum_{i,j \in U} \frac{\epsilon(e^t_i) \epsilon(e^s_j)}{\epsilon\rvert_{U}(\op{1})} \Lambda_{Uij}.
\eeq	
Here, $U$ label the irreducible representations of the target counital subalgebra, $\{e^t_i\}$ and $\{e^s_i\}$ are dual orthonormal bases for $H_t$ and $H_s$, and $\Lambda_{Uij}$ are matrix units which specify a block decomposition of $H$ into the direct sum of ordinary Hopf algebras
\beq
	H \simeq \bigoplus_{Uij} H_{Uij}.
\eeq 
Given $\Lambda$, the projection becomes
\begin{flalign} \label{WHA Projection}
	&E: S \rightarrow S^H, \nonumber \\ 
	&E(s) = \rho_{\Lambda}(s) = \sum_{U \in \textrm{IrrRep}(H_t)} \sum_{i,j \in U} \frac{\epsilon(e^t_i) \epsilon(e^s_j)}{\epsilon\rvert_{U}(\op{1})} \rho_{\Lambda_{Uij}}(s), 
\end{flalign}
where $\rho_{\Lambda_{Uij}}(s)$ can be interpreted immediately as the invariantizing projection for the action of the sector $H_{Uij} \subset H$. 
 
In the case that $H = L^1(\mathcal{G})$ encodes the action of a groupoid, each irrep is one-dimensional and the prefactor $\frac{\epsilon(e^t_i) \epsilon(e^s_j)}{\epsilon\rvert_{U}(\op{1})}$ reduces to a weighting assigned to each $\alpha$-sector. The matrix unit $\Lambda_{\alpha}$ corresponds to the Haar element of the group $G_{\alpha}$ which takes the form of an average over the group:
\beq
	\Lambda_{\alpha} = \frac{1}{|G_{\alpha}|} \sum_{g \in G_{\alpha}} \ell_{g}. 
\eeq	
Thus, in this case, the invariantizing projection takes the form
\beq
	E(s) = \int_{X} d\mu(\alpha) \; \frac{1}{|G_{\alpha}|} \sum_{g \in G_{\alpha}} \rho_g(s) = \int_{X} d\mu(\alpha) \; s_{\alpha}. 
\eeq
reproducing the formula \eqref{Non Mult Fact}. 

\begin{comment}
For the partition function, we find
\beq
	\mathcal{Z}(M_1 \sqcup ... \sqcup M_n) = E\big(\mathcal{F}(M_1 \sqcup ... \sqcup M_n)\big) = \int_{X} d\mu(\alpha) \; Z_{\mathcal{F}}^{\alpha}(M_1) \cdot ... \cdot Z_{\mathcal{F}}^{\alpha}(M_n),
\eeq
which reduces to $Z_{\mathcal{F}}$ in the event that there is exactly one $\alpha$-sector e.g. when multiplicativity is restored. 
\end{comment}

In the gravitational context, the projection \eqref{WHA Projection} thereby combines both the averaging over $\alpha$-sectors and the averaging over $\mu$-sectors which results in the universal construction starting from the unitary QFT into which it embeds \cite{McNamara:2026isz}. Rather than interpret eqn. \eqref{WHA Projection} as an averaging over an ensemble of unitary theories, we have now argued that it can be derived as the application of an invariantizing projection in a \emph{single} unitary QFT. To go from the incomplete, universal theory to the extended, unitary QFT one simply adds back the $\alpha$-sectors and the half ER bridges, both of which are coherently encoded within a single algebraic object in the form of the weak Hopf algebra $H$ reconstructed by the baby universe category. 

We can consider this analysis in the more general case in which a theory $\mathcal{U}$ fails to factorize due to an incompleteness in the spectrum of a \emph{non-invertible} symmetry encoded in a weak Hopf algebra $H$. This is accomplished by appealing to Lemma \ref{lemma: Inv CP}, so that for each $B \in \textrm{Bord}_d$ we have $\mathscr{A}_{\mathcal{F}}(B) \simeq \mathscr{A}_{\mathcal{U}}(B) \times_{\hat{\rho}} \hat{H}$. 
%The Hilbert spaces associated to $\mathcal{F}$ are of the form $\mathcal{U}_{\mathcal{Z}}(B) \otimes \mathscr{H}_{\textrm{BU}}$, where the baby universe Hilbert space is isomorphic to the GNS Hilbert space of $H_t$, reducing to $\mathbb{C}$ in the multiplicative case. 
Working in the case in which $H$ is allowed to be a general weak Hopf algebra leaves open the possibility that the baby-universe algebra $H_t$ may be noncommutative. We can realize a standard representation of $\mathscr{A}_{\mathcal{F}}(B)$ acting on its GNS Hilbert space, which is of the form\footnote{The tensor product notation is the second equality is a bit subtle. More concretely, we should think of the space of ER bridges as being \emph{fibered} over the space of BU states, as is described in the discussion following eqn. \eqref{GNS HS for full alg}.}
\begin{flalign} \label{GNS HS for full alg}
	\mathscr{H}_{\textrm{GNS}}(\mathscr{A}_{\mathcal{F}}) &= \mathscr{H}_{\textrm{GNS}}(\mathscr{A}_{\mathcal{Z}}) \otimes \mathscr{H}_{\textrm{GNS}}(H) \nonumber \\
	&\simeq \mathscr{H}_{\textrm{GNS}}(\mathscr{A}_{\mathcal{Z}}) \otimes \mathscr{H}_{\textrm{BU}} \otimes \mathscr{H}_{\textrm{ER}}. 
\end{flalign}
The presence of the counital target/source subalgebra implies that $\mathscr{H}_{\textrm{GNS}}(H)$ admits a dense set of states of the form $\ket{U,i,j,h}$ where $(U,i,j)$ describes a `generalized $\alpha$-sector' and $h \in H_{Uij}$ is a `generalized ER-bridge' fibered over this sector -- hence the decomposition in \eqref{GNS HS for full alg}. Here, $\mathscr{H}_{\textrm{BU}}$ can be read as the GNS Hilbert space of the target/source counital algebra, carrying the generalized $\alpha$-sectors, while $\mathscr{H}_{\textrm{ER}}$ is the Hilbert space of (half) ER bridges.\footnote{Notice, if the partition function associated with the theory $\mathcal{U}$ factorizes, the unitary QFT it embeds into will be irreducible. By consequence, we find $H_t = \mathbb{C}$ and thus $\mathscr{H}_{\textrm{BU}} \simeq \mathbb{C}$, as expected.} 

The generalized $\alpha$-sectors fit into a single theory rather than defining distinct theories within an ensemble. The map \eqref{WHA Projection} may therefore be regarded as a general form of the filter described in Definition \ref{Def Filter}, with the unitary quantum field theory $\mathcal{F}$ its associated extension. More generally, we expect the above algebraic structure to appear whenever a system admits a `weak' non-invertible symmetry \cite{AliAhmad:2025bnd}.

% ------------------------------------------------------------------
% Acknowledgments
% ------------------------------------------------------------------

\begin{acknowledgments}
It is our pleasure to thank Shadi Ali Ahmad, Sulaiman Alvi, Vijay Balasubramanian, Charlie Cummings, Laurent Friedel, Nima Lashkari, Hong Liu, Jake McNamara, Daniel Murphy, Erik Verlinde, Zhencheng Wang, and Tom Yildirim for many helpful discussions. The work of M.S.K. was supported by the Heising-Simons foundation ``Observable Signatures of Quantum Gravity" collaboration and the Walter Burke Institute for Theoretical Physics. This material is also based upon work supported by the U.S. Department of Energy, Office of Science, Office of High Energy Physics, under Award Number DE-SC0011632. 
\end{acknowledgments}

% ------------------------------------------------------------------
% Bibliography
% ------------------------------------------------------------------

\bibliographystyle{uiuchept}
\bibliography{arXivSubFilter}

@article{Witten:1999xp,
    author = "Witten, Edward and Yau, Shing-Tung",
    editor = "D'Hoker, Erik and Phong, Duong and Yau, Shing-Tung",
    title = "{Connectedness of the boundary in the AdS / CFT correspondence}",
    eprint = "hep-th/9910245",
    archivePrefix = "arXiv",
    doi = "10.4310/ATMP.1999.v3.n6.a1",
    journal = "Adv. Theor. Math. Phys.",
    volume = "3",
    pages = "1635--1655",
    year = "1999"
}

@article{Maldacena:2004rf,
    author = "Maldacena, Juan Martin and Maoz, Liat",
    title = "{Wormholes in AdS}",
    eprint = "hep-th/0401024",
    archivePrefix = "arXiv",
    reportNumber = "ITFA-2003-57",
    doi = "10.1088/1126-6708/2004/02/053",
    journal = "JHEP",
    volume = "02",
    pages = "053",
    year = "2004"
}

@article{Coleman:1988cy,
    author = "Coleman, Sidney R.",
    title = "{Black holes as red herrings: Topological fluctuations and the loss of quantum coherence}",
    reportNumber = "HUTP-88/A008",
    doi = "10.1016/0550-3213(88)90110-1",
    journal = "Nucl. Phys. B",
    volume = "307",
    pages = "867--882",
    year = "1988"
}

@article{Giddings:1987cg,
    author = "Giddings, Steven B. and Strominger, Andrew",
    title = "{Axion Induced Topology Change in Quantum Gravity and String Theory}",
    reportNumber = "HUTP-87-A067",
    doi = "10.1016/0550-3213(88)90446-4",
    journal = "Nucl. Phys. B",
    volume = "306",
    pages = "890--907",
    year = "1988"
}

@article{Giddings:1988cx,
    author = "Giddings, Steven B. and Strominger, Andrew",
    title = "{Loss of incoherence and determination of coupling constants in quantum gravity}",
    reportNumber = "HUTP-88/A006",
    doi = "10.1016/0550-3213(88)90109-5",
    journal = "Nucl. Phys. B",
    volume = "307",
    pages = "854--866",
    year = "1988"
}

@article{Hawking:1976ra,
    author = "Hawking, S. W.",
    title = "{Breakdown of Predictability in Gravitational Collapse}",
    doi = "10.1103/PhysRevD.14.2460",
    journal = "Phys. Rev. D",
    volume = "14",
    pages = "2460--2473",
    year = "1976"
}

@article{Hawking:1975vcx,
    author = "Hawking, S. W.",
    editor = "Gibbons, G. W. and Hawking, S. W.",
    title = "{Particle Creation by Black Holes}",
    doi = "10.1007/BF02345020",
    journal = "Commun. Math. Phys.",
    volume = "43",
    pages = "199--220",
    year = "1975",
    note = "[Erratum: Commun.Math.Phys. 46, 206 (1976)]"
}

@inproceedings{Page:1993up,
    author = "Page, Don N.",
    title = "{Black hole information}",
    booktitle = "{5th Canadian Conference on General Relativity and Relativistic Astrophysics (5CCGRRA)}",
    eprint = "hep-th/9305040",
    archivePrefix = "arXiv",
    reportNumber = "ALBERTA-THY-23-93",
    month = "5",
    year = "1993"
}

@inproceedings{Polchinski:2016hrw,
    author = "Polchinski, Joseph",
    title = "{The black hole information problem.}",
    booktitle = "{Theoretical Advanced Study Institute in Elementary Particle Physics}: {New Frontiers in Fields and Strings}",
    eprint = "1609.04036",
    archivePrefix = "arXiv",
    primaryClass = "hep-th",
    doi = "10.1142/9789813149441_0006",
    pages = "353--397",
    year = "2017"
}

@article{Penington:2019kki,
    author = "Penington, Geoff and Shenker, Stephen H. and Stanford, Douglas and Yang, Zhenbin",
    title = "{Replica wormholes and the black hole interior}",
    eprint = "1911.11977",
    archivePrefix = "arXiv",
    primaryClass = "hep-th",
    doi = "10.1007/JHEP03(2022)205",
    journal = "JHEP",
    volume = "03",
    pages = "205",
    year = "2022"
}

@article{Almheiri:2019qdq,
    author = "Almheiri, Ahmed and Hartman, Thomas and Maldacena, Juan and Shaghoulian, Edgar and Tajdini, Amirhossein",
    title = "{Replica Wormholes and the Entropy of Hawking Radiation}",
    eprint = "1911.12333",
    archivePrefix = "arXiv",
    primaryClass = "hep-th",
    doi = "10.1007/JHEP05(2020)013",
    journal = "JHEP",
    volume = "05",
    pages = "013",
    year = "2020"
}

@article{Liu:2025ikq,
    author = "Liu, Hong",
    title = "{''Filtering'' CFTs at large N: Euclidean Wormholes, Closed Universes, and Black Hole Interiors}",
    eprint = "2512.13807",
    archivePrefix = "arXiv",
    primaryClass = "hep-th",
    reportNumber = "MIT-CTP/5970",
    month = "12",
    year = "2025"
}

@article{Liu:2025cml,
    author = "Liu, Hong",
    title = "{Towards a holographic description of closed universes}",
    eprint = "2509.14327",
    archivePrefix = "arXiv",
    primaryClass = "hep-th",
    reportNumber = "MIT-CTP/5923",
    month = "9",
    year = "2025"
}

@article{Liu:2026fnd,
    author = "Liu, Hong",
    title = "{Ramp, Plateau, and Wormholes without Averaging, and Hyper-non-perturbative Structures in Gravity}",
    eprint = "2608.02743",
    archivePrefix = "arXiv",
    primaryClass = "hep-th",
    reportNumber = "MIT-CTP/6074",
    month = "8",
    year = "2026"
}

@article{Kudler-Flam:2025cki,
    author = "Kudler-Flam, Jonah and Witten, Edward",
    title = "{Emergent mixed states for baby universes and black holes}",
    eprint = "2510.06376",
    archivePrefix = "arXiv",
    primaryClass = "hep-th",
    doi = "10.1007/JHEP05(2026)090",
    journal = "JHEP",
    volume = "05",
    pages = "090",
    year = "2026"
}

@article{Kudler-Flam:2026nzz,
    author = "Kudler-Flam, Jonah and Witten, Edward",
    title = "{Wormholes and Averaging over N}",
    eprint = "2605.15180",
    archivePrefix = "arXiv",
    primaryClass = "hep-th",
    month = "5",
    year = "2026"
}

@article{Klinger:2025tvg,
    author = "Klinger, Marc",
    title = "{A Theory of Backgrounds and Background Independence}",
    eprint = "2512.05043",
    archivePrefix = "arXiv",
    primaryClass = "hep-th",
    month = "12",
    year = "2025"
}

@article{Klinger:2026kqj,
    author = "Klinger, Marc S.",
    title = "{How to have your wormholes and factorize, too}",
    eprint = "2602.15120",
    archivePrefix = "arXiv",
    primaryClass = "hep-th",
    month = "2",
    year = "2026"
}

@article{Gesteau:2025obm,
    author = "Gesteau, Elliott",
    title = "{A no-go theorem for large $N$ closed universes}",
    eprint = "2509.14338",
    archivePrefix = "arXiv",
    primaryClass = "hep-th",
    reportNumber = "MIT-CTP/5927",
    month = "9",
    year = "2025"
}

@article{McNamara:2020uza,
    author = "McNamara, Jacob and Vafa, Cumrun",
    title = "{Baby Universes, Holography, and the Swampland}",
    eprint = "2004.06738",
    archivePrefix = "arXiv",
    primaryClass = "hep-th",
    month = "4",
    year = "2020"
}

@article{Harlow:2026hky,
    author = "Harlow, Daniel",
    title = "{Observers, $\alpha$-parameters, and the Hartle-Hawking state}",
    eprint = "2602.03835",
    archivePrefix = "arXiv",
    primaryClass = "hep-th",
    reportNumber = "MIT-CTP/5900",
    month = "2",
    year = "2026"
}

@article{Antonini:2024mci,
    author = "Antonini, Stefano and Rath, Pratik",
    title = "{Do holographic CFT states have unique semiclassical bulk duals?}",
    eprint = "2408.02720",
    archivePrefix = "arXiv",
    primaryClass = "hep-th",
    doi = "10.1142/S0218271825440250",
    journal = "Int. J. Mod. Phys. D",
    volume = "34",
    number = "16",
    pages = "2544025",
    year = "2025"
}

@article{Harlow:2025pvj,
    author = "Harlow, Daniel and Usatyuk, Mykhaylo and Zhao, Ying",
    title = "{Quantum mechanics and observers for gravity in a closed universe}",
    eprint = "2501.02359",
    archivePrefix = "arXiv",
    primaryClass = "hep-th",
    reportNumber = "MIT-CTP/5824",
    doi = "10.1007/JHEP02(2026)108",
    journal = "JHEP",
    volume = "02",
    pages = "108",
    year = "2026"
}

@article{Antonini:2025ioh,
    author = "Antonini, Stefano and Rath, Pratik and Sasieta, Martin and Swingle, Brian and Vilar L{\'o}pez, Alejandro",
    title = "{The baby universe is fine and the CFT knows it: on holography for closed universes}",
    eprint = "2507.10649",
    archivePrefix = "arXiv",
    primaryClass = "hep-th",
    doi = "10.1007/JHEP12(2025)159",
    journal = "JHEP",
    volume = "12",
    pages = "159",
    year = "2025"
}

@article{Sasieta:2025vck,
    author = "Sasieta, Martin and Swingle, Brian and Vilar L{\'o}pez, Alejandro",
    title = "{Baby Universes from Thermal Pure States in the Sachdev-Ye-Kitaev Model}",
    eprint = "2512.00149",
    archivePrefix = "arXiv",
    primaryClass = "hep-th",
    doi = "10.1103/8p6x-w737",
    journal = "Phys. Rev. Lett.",
    volume = "137",
    number = "4",
    pages = "041501",
    year = "2026"
}

@article{Higginbotham:2025clp,
    author = "Higginbotham, Kenneth",
    title = "{Helping observers in closed universes reach their full potential}",
    eprint = "2512.17993",
    archivePrefix = "arXiv",
    primaryClass = "hep-th",
    doi = "10.1007/JHEP03(2026)183",
    journal = "JHEP",
    volume = "03",
    pages = "183",
    year = "2026"
}

@article{Giddings:1988wv,
    author = "Giddings, Steven B. and Strominger, Andrew",
    title = "{Baby Universes, Third Quantization and the Cosmological Constant}",
    reportNumber = "HUTP-88/A036",
    doi = "10.1016/0550-3213(89)90353-2",
    journal = "Nucl. Phys. B",
    volume = "321",
    pages = "481--508",
    year = "1989"
}

@article{VanRaamsdonk:2026tnv,
    author = "Van Raamsdonk, Mark and Vilar L{\'o}pez, Alejandro",
    title = "{Menagerie of Euclidean constructions for 3D holographic cosmologies}",
    eprint = "2601.10906",
    archivePrefix = "arXiv",
    primaryClass = "hep-th",
    doi = "10.1103/qlnl-rpm2",
    journal = "Phys. Rev. D",
    volume = "113",
    number = "10",
    pages = "106010",
    year = "2026"
}

@article{Abdalla:2025gzn,
    author = "Abdalla, Ahmed I. and Antonini, Stefano and Iliesiu, Luca V. and Levine, Adam",
    title = "{The gravitational path integral from an observer{\textquoteright}s point of view}",
    eprint = "2501.02632",
    archivePrefix = "arXiv",
    primaryClass = "hep-th",
    doi = "10.1007/JHEP05(2025)059",
    journal = "JHEP",
    volume = "05",
    pages = "059",
    year = "2025"
}

@article{Antonini:2023hdh,
    author = "Antonini, Stefano and Sasieta, Martin and Swingle, Brian",
    title = "{Cosmology from random entanglement}",
    eprint = "2307.14416",
    archivePrefix = "arXiv",
    primaryClass = "hep-th",
    doi = "10.1007/JHEP11(2023)188",
    journal = "JHEP",
    volume = "11",
    pages = "188",
    year = "2023"
}

@article{Saad:2021rcu,
    author = "Saad, Phil and Shenker, Stephen H. and Stanford, Douglas and Yao, Shunyu",
    title = "{Wormholes without averaging}",
    eprint = "2103.16754",
    archivePrefix = "arXiv",
    primaryClass = "hep-th",
    doi = "10.1007/JHEP09(2024)133",
    journal = "JHEP",
    volume = "09",
    pages = "133",
    year = "2024"
}

@article{Mukhametzhanov:2021nea,
    author = "Mukhametzhanov, Baur",
    title = "{Half-wormholes in SYK with one time point}",
    eprint = "2105.08207",
    archivePrefix = "arXiv",
    primaryClass = "hep-th",
    doi = "10.21468/SciPostPhys.12.1.029",
    journal = "SciPost Phys.",
    volume = "12",
    number = "1",
    pages = "029",
    year = "2022"
}

@article{Yang:2025kgs,
    author = "Yang, Yingyu",
    title = "{Half-wormholes in a complex SYK model}",
    eprint = "2503.13172",
    archivePrefix = "arXiv",
    primaryClass = "hep-th",
    doi = "10.1209/0295-5075/adfd7d",
    journal = "EPL",
    volume = "151",
    number = "5",
    pages = "59001",
    year = "2025"
}

@article{Blommaert:2019wfy,
    author = "Blommaert, Andreas and Mertens, Thomas G. and Verschelde, Henri",
    title = "{Eigenbranes in Jackiw-Teitelboim gravity}",
    eprint = "1911.11603",
    archivePrefix = "arXiv",
    primaryClass = "hep-th",
    doi = "10.1007/JHEP02(2021)168",
    journal = "JHEP",
    volume = "02",
    pages = "168",
    year = "2021"
}

@article{Blommaert:2021gha,
    author = "Blommaert, Andreas and Kruthoff, Jorrit",
    title = "{Gravity without averaging}",
    eprint = "2107.02178",
    archivePrefix = "arXiv",
    primaryClass = "hep-th",
    doi = "10.21468/SciPostPhys.12.2.073",
    journal = "SciPost Phys.",
    volume = "12",
    number = "2",
    pages = "073",
    year = "2022"
}

@article{Garcia-Garcia:2021squ,
    author = "Garc{\'\i}a-Garc{\'\i}a, Antonio M. and Godet, Victor",
    title = "{Half-wormholes in nearly AdS$_2$ holography}",
    eprint = "2107.07720",
    archivePrefix = "arXiv",
    primaryClass = "hep-th",
    doi = "10.21468/SciPostPhys.12.4.135",
    journal = "SciPost Phys.",
    volume = "12",
    number = "4",
    pages = "135",
    year = "2022"
}

@article{DiUbaldo:2023qli,
    author = "Di Ubaldo, Gabriele and Perlmutter, Eric",
    title = "{AdS$_{3}$/RMT$_{2}$ duality}",
    eprint = "2307.03707",
    archivePrefix = "arXiv",
    primaryClass = "hep-th",
    doi = "10.1007/JHEP12(2023)179",
    journal = "JHEP",
    volume = "12",
    pages = "179",
    year = "2023"
}

@article{McNamara:2026isz,
    author = "McNamara, Jacob and Wang, Zhencheng",
    title = "{Wormholes as red herrings: reflection positivity and the reconstruction of unitary quantum field theories}",
    eprint = "2607.01322",
    archivePrefix = "arXiv",
    primaryClass = "hep-th",
    month = "7",
    year = "2026"
}

@article{BLANCHET1995883,
title = {Topological Quantum Field Theories derived from the Kauffman bracket},
journal = {Topology},
volume = {34},
number = {4},
pages = {883-927},
year = {1995},
issn = {0040-9383},
doi = {https://doi.org/10.1016/0040-9383(94)00051-4},
url = {https://www.sciencedirect.com/science/article/pii/0040938394000514},
author = {C. Blanchet and N. Habegger and G. Masbaum and P. Vogel}
}

@article{Colafranceschi:2023moh,
    author = "Colafranceschi, Eugenia and Dong, Xi and Marolf, Donald and Wang, Zhencheng",
    title = "{Algebras and Hilbert spaces from gravitational path integrals. Understanding Ryu-Takayanagi/HRT as entropy without AdS/CFT}",
    eprint = "2310.02189",
    archivePrefix = "arXiv",
    primaryClass = "hep-th",
    doi = "10.1007/JHEP10(2024)063",
    journal = "JHEP",
    volume = "10",
    pages = "063",
    year = "2024"
}

@article{bunke2025additiveccategoriesktheory,
      title={Additive C*-categories and K-theory}, 
      author={Ulrich Bunke and Alexander Engel},
      year={2025},
      eprint={2010.14830},
      archivePrefix={arXiv},
      primaryClass={math.KT},
      url={https://arxiv.org/abs/2010.14830}, 
}

@article{GhezLimaRoberts1985,
  author  = {Ghez, P. and Lima, R. and Roberts, J. E.},
  title   = {$W^*$-categories},
  journal = {Pacific Journal of Mathematics},
  volume  = {120},
  number  = {1},
  pages   = {79--109},
  year    = {1985},
  doi     = {10.2140/pjm.1985.120.79}
}

@article{henriques2024completewcategories,
      title={Complete W*-categories}, 
      author={André Henriques and Nivedita and David Penneys},
      year={2024},
      eprint={2411.01678},
      archivePrefix={arXiv},
      primaryClass={math.OA},
      url={https://arxiv.org/abs/2411.01678}, 
}

@article{Binder:2019zqc,
    author = "Binder, Damon J. and Rychkov, Slava",
    title = "{Deligne Categories in Lattice Models and Quantum Field Theory, or Making Sense of $O(N)$ Symmetry with Non-integer $N$}",
    eprint = "1911.07895",
    archivePrefix = "arXiv",
    primaryClass = "hep-th",
    reportNumber = "PUPT-2601",
    doi = "10.1007/JHEP04(2020)117",
    journal = "JHEP",
    volume = "04",
    pages = "117",
    year = "2020"
}

@article{Marolf:2020xie,
    author = "Marolf, Donald and Maxfield, Henry",
    title = "{Transcending the ensemble: baby universes, spacetime wormholes, and the order and disorder of black hole information}",
    eprint = "2002.08950",
    archivePrefix = "arXiv",
    primaryClass = "hep-th",
    doi = "10.1007/JHEP08(2020)044",
    journal = "JHEP",
    volume = "08",
    pages = "044",
    year = "2020"
}

@article{Balasubramanian:2025jeu,
    author = "Balasubramanian, Vijay and Yildirim, Tom",
    title = "{Nonperturbative toolkit for quantum gravity}",
    eprint = "2504.16986",
    archivePrefix = "arXiv",
    primaryClass = "hep-th",
    doi = "10.1103/nlzj-w34h",
    journal = "Phys. Rev. D",
    volume = "114",
    number = "2",
    pages = "026035",
    year = "2026"
}

@article{Balasubramanian:2025hns,
    author = "Balasubramanian, Vijay and Yildirim, Tom",
    title = "{How to count states in gravity}",
    eprint = "2506.15767",
    archivePrefix = "arXiv",
    primaryClass = "hep-th",
    doi = "10.1103/9v3t-91qh",
    journal = "Phys. Rev. D",
    volume = "114",
    number = "2",
    pages = "026030",
    year = "2026"
}

@article{Balasubramanian:2025akx,
    author = "Balasubramanian, Vijay and Yildirim, Tom",
    title = "{Observing spacetime}",
    eprint = "2509.09763",
    archivePrefix = "arXiv",
    primaryClass = "hep-th",
    doi = "10.1103/p7wt-47rj",
    journal = "Phys. Rev. D",
    volume = "113",
    number = "10",
    pages = "106034",
    year = "2026"
}

@article{AliAhmad:2025bnd,
    author = "Ali Ahmad, Shadi and Klinger, Marc S. and Wang, Yifan",
    title = "{The many faces of non-invertible symmetries}",
    eprint = "2509.18072",
    archivePrefix = "arXiv",
    primaryClass = "hep-th",
    doi = "10.1007/JHEP05(2026)110",
    journal = "JHEP",
    volume = "05",
    pages = "110",
    year = "2026"
}

@article{AliAhmad:2025oli,
    author = "Ali Ahmad, Shadi and Klinger, Marc S.",
    title = "{Extensions from within}",
    eprint = "2503.02944",
    archivePrefix = "arXiv",
    primaryClass = "hep-th",
    month = "3",
    year = "2025"
}

@article{Doplicher:1971wk,
    author = "Doplicher, Sergio and Haag, Rudolf and Roberts, John E.",
    title = "{Local observables and particle statistics. 1}",
    doi = "10.1007/BF01877742",
    journal = "Commun. Math. Phys.",
    volume = "23",
    pages = "199--230",
    year = "1971"
}

@article{Doplicher:1973at,
    author = "Doplicher, Sergio and Haag, Rudolf and Roberts, John E.",
    title = "{Local observables and particle statistics. 2}",
    doi = "10.1007/BF01646454",
    journal = "Commun. Math. Phys.",
    volume = "35",
    pages = "49--85",
    year = "1974"
}

@article{Doplicher:1989afv,
    author = "Doplicher, Sergio and Roberts, John E.",
    title = "{A new duality theory for compact groups}",
    doi = "10.1007/BF01388849",
    journal = "Invent. Math.",
    volume = "98",
    number = "1",
    pages = "157--218",
    year = "1989"
}

@article{Ostrik:2001xnt,
    author = "Ostrik, Viktor",
    title = "{Module categories, weak Hopf algebras and modular invariants}",
    eprint = "math/0111139",
    archivePrefix = "arXiv",
    doi = "10.1007/s00031-003-0515-6",
    journal = "Transform. Groups",
    volume = "8",
    number = "2",
    pages = "177--206",
    year = "2003"
}

@inproceedings{Rehren:1996ku,
    author = "Rehren, Karl-Henning",
    title = "{Weak C* Hopf symmetry}",
    booktitle = "{21st International Colloquium on Group Theoretical Methods in Physics}",
    eprint = "q-alg/9611007",
    archivePrefix = "arXiv",
    reportNumber = "DESY-96-231",
    pages = "62--69",
    month = "11",
    year = "1996"
}

@article{Etingof:2002vpd,
    author = "Etingof, Pavel and Nikshych, Dmitri and Ostrik, Viktor",
    title = "{On fusion categories}",
    eprint = "math/0203060",
    archivePrefix = "arXiv",
    month = "3",
    year = "2002"
}

@article{Bai:2025zze,
    author = "Bai, Ansi and Zhang, Zhi-Hao",
    title = "{On the Representation Categories of Weak Hopf Algebras Arising from Levin-Wen Models}",
    eprint = "2503.06731",
    archivePrefix = "arXiv",
    primaryClass = "math.QA",
    month = "3",
    year = "2025"
}

@article{Jones:1983kv,
    author = "Jones, V. F. R.",
    title = "{Index for subfactors}",
    doi = "10.1007/BF01389127",
    journal = "Invent. Math.",
    volume = "72",
    pages = "1--25",
    year = "1983"
}

@article{PimsnerPopa1986,
  author    = {Pimsner, Mihai and Popa, Sorin},
  title     = {Entropy and index for subfactors},
  journal   = {Annales scientifiques de l'École Normale Supérieure},
  series    = {4},
  volume    = {19},
  number    = {1},
  pages     = {57--106},
  year      = {1986},
  doi       = {10.24033/asens.1504},
  url       = {https://www.numdam.org/articles/10.24033/asens.1504/}
}

@Inbook{Kosaki1991,
author="Kosaki, Hideki",
editor="Araki, Huzihiro
and Kadison, Richard V.",
title="Index Theory for Type III Factors",
bookTitle="Mappings of Operator Algebras: Proceedings of the Japan---U.S. Joint Seminar, University of Pennsylvania, 1988",
year="1991",
publisher="Birkh{\"a}user Boston",
address="Boston, MA",
pages="227--231",
isbn="978-1-4612-0453-4",
doi="10.1007/978-1-4612-0453-4_11",
url="https://doi.org/10.1007/978-1-4612-0453-4_11"
}

@book{Watatani1990,
  author    = {Yasuo Watatani},
  title     = {Index for {$C^*$}-Subalgebras},
  series    = {Memoirs of the American Mathematical Society},
  volume    = {83},
  number    = {424},
  publisher = {American Mathematical Society},
  address   = {Providence, RI},
  year      = {1990},
  doi       = {10.1090/memo/0424},
  isbn      = {9780821824870}
}

@article{ENOCK1996466,
title = {Irreducible Inclusions of Factors, Multiplicative Unitaries, and Kac Algebras},
journal = {Journal of Functional Analysis},
volume = {137},
number = {2},
pages = {466-543},
year = {1996},
issn = {0022-1236},
doi = {https://doi.org/10.1006/jfan.1996.0053},
url = {https://www.sciencedirect.com/science/article/pii/S0022123696900531},
author = {Michel Enock and Ryszard Nest}
}

@article{Bischoff:2014xea,
    author = "Bischoff, Marcel and Longo, Roberto and Kawahigashi, Yasuyuki and Rehren, Karl-Henning",
    title = "{Tensor categories and endomorphisms of von Neumann algebras (with applications to Quantum Field Theory)}",
    eprint = "1407.4793",
    archivePrefix = "arXiv",
    primaryClass = "math.OA",
    doi = "10.1007/978-3-319-14301-9",
    journal = "Physics",
    volume = "3",
    pages = "1--94",
    year = "2015"
}

@article{Longo:1994zza,
    author = "Longo, Roberto",
    title = "{A duality for Hopf algebras and for subfactors. 1.}",
    doi = "10.1007/BF02100488",
    journal = "Commun. Math. Phys.",
    volume = "159",
    pages = "133--150",
    year = "1994"
}

@article{Longo:1989tt,
    author = "Longo, R.",
    title = "{Index of subfactors and statistics of quantum fields. I}",
    doi = "10.1007/BF02125124",
    journal = "Commun. Math. Phys.",
    volume = "126",
    pages = "217--247",
    year = "1989"
}

@article{DelVecchio:2017axj,
    author = "Del Vecchio, Simone and Giorgetti, Luca",
    title = "{Infinite index extensions of local nets and defects}",
    eprint = "1703.03605",
    archivePrefix = "arXiv",
    primaryClass = "math.OA",
    doi = "10.1142/S0129055X18500022",
    journal = "Rev. Math. Phys.",
    volume = "30",
    number = "02",
    pages = "1850002",
    year = "2018"
}

@article{nikshych2002finite,
  title={Finite quantum groupoids and their applications},
  author={Nikshych, Dmitri and Vainerman, Leonid},
  journal={New directions in Hopf algebras},
  volume={43},
  pages={211--262},
  year={2002}
}

@article{nikshych1999dualitytheoremquantumgroupoids,
      title={A Duality Theorem for Quantum Groupoids}, 
      author={Dmitri Nikshych},
      year={1999},
      eprint={math/9912226},
      archivePrefix={arXiv},
      primaryClass={math.QA},
      url={https://arxiv.org/abs/math/9912226}, 
}

@article{BOHM1999385,
title = {Weak Hopf Algebras: I. Integral Theory and C-Structure},
journal = {Journal of Algebra},
volume = {221},
number = {2},
pages = {385-438},
year = {1999},
issn = {0021-8693},
doi = {https://doi.org/10.1006/jabr.1999.7984},
url = {https://www.sciencedirect.com/science/article/pii/S002186939997984X},
author = {Gabriella Böhm and Florian Nill and Kornél Szlachányi}
}

@article{bohm1999weakhopfalgebrasii,
      title={Weak Hopf Algebras II: Representation theory, dimensions and the Markov trace}, 
      author={G. Bohm and K. Szlachanyi},
      year={1999},
      eprint={math/9906045},
      archivePrefix={arXiv},
      primaryClass={math.QA},
      url={https://arxiv.org/abs/math/9906045}, 
}

@article{Nill:1998iw,
    author = "Nill, F. and Szlachanyi, K. and Wiesbrock, H. W.",
    title = "{Weak Hopf algebras and reducible Jones inclusions of depth 2.: 1. From crossed products to Jones towers}",
    reportNumber = "SFB-288-335",
    month = "8",
    year = "1998"
}

@article{Donnelly:2016auv,
    author = "Donnelly, William and Freidel, Laurent",
    title = "{Local subsystems in gauge theory and gravity}",
    eprint = "1601.04744",
    archivePrefix = "arXiv",
    primaryClass = "hep-th",
    doi = "10.1007/JHEP09(2016)102",
    journal = "JHEP",
    volume = "09",
    pages = "102",
    year = "2016"
}

@article{Klinger:2023tgi,
    author = "Klinger, Marc S. and Leigh, Robert G.",
    title = "{Crossed products, extended phase spaces and the resolution of entanglement singularities}",
    eprint = "2306.09314",
    archivePrefix = "arXiv",
    primaryClass = "hep-th",
    doi = "10.1016/j.nuclphysb.2024.116453",
    journal = "Nucl. Phys. B",
    volume = "999",
    pages = "116453",
    year = "2024"
}

@article{Ciambelli:2026vxa,
    author = "Ciambelli, Luca and Klinger, Marc S.",
    title = "{Quantization of Gravity on Null Hypersurfaces}",
    eprint = "2607.07785",
    archivePrefix = "arXiv",
    primaryClass = "hep-th",
    month = "7",
    year = "2026"
}

@article{nikshych2000galoiscorrespondenceii1factors,
      title="{A Galois correspondence for II$_1$ factors and quantum groupoids}", 
      author={Dmitri Nikshych and Leonid Vainerman},
      year={2000},
      eprint={math/0001020},
      archivePrefix={arXiv},
      primaryClass={math.QA},
      url={https://arxiv.org/abs/math/0001020}, 
}

@article{Marolf:2024adj,
    author = "Marolf, Donald and Zhang, Daiming",
    title = "{When left and right disagree: entropy and von Neumann algebras in quantum gravity with general AlAdS boundary conditions}",
    eprint = "2402.09691",
    archivePrefix = "arXiv",
    primaryClass = "hep-th",
    doi = "10.1007/JHEP08(2024)010",
    journal = "JHEP",
    volume = "08",
    pages = "010",
    year = "2024"
}

@article{Freed:2012bs,
    author = "Freed, Daniel S. and Teleman, Constantin",
    title = "{Relative quantum field theory}",
    eprint = "1212.1692",
    archivePrefix = "arXiv",
    primaryClass = "hep-th",
    doi = "10.1007/s00220-013-1880-1",
    journal = "Commun. Math. Phys.",
    volume = "326",
    pages = "459--476",
    year = "2014"
}

\end{document}